\documentclass[superscriptaddress,
 amsmath,amssymb,
 aps,
prl,
reprint
]{revtex4-2}

\usepackage{qcircuit}
\usepackage[dvips]{graphicx}
\usepackage{amsmath,amssymb,amsthm,mathrsfs,amsfonts,dsfont}
\usepackage{mathtools}
\usepackage[caption=false]{subfig}
\usepackage{braket}
\usepackage{bm}
\usepackage{enumerate}
\usepackage{xcolor}
\usepackage{comment}
\usepackage{scalefnt}
\usepackage[colorlinks = true, 
citecolor = blue,
linkcolor = blue,
urlcolor = blue]{hyperref}
\usepackage{makecell}

\usepackage{relsize}
\usepackage{pagecolor}
\usepackage[T1]{fontenc}
\usepackage{lmodern}
\usepackage[utf8]{inputenc}
\usepackage{microtype}

\usepackage{tikzscale}
\usepackage{pgfplots}
\pgfplotsset{compat=newest}
\usetikzlibrary{plotmarks}
\usetikzlibrary{arrows.meta}
\usetikzlibrary{external}
\usepgfplotslibrary{patchplots}
\usepackage{grffile}
\usepackage{enumitem}

\DeclareMathOperator{\Tr}{Tr}

\DeclareMathSymbol{\mhyphen}{\mathord}{AMSa}{"39}

\newcommand{\coleq}{\mathrel{\mathop:}\nobreak\mkern-1.2mu=}

\newcommand{\mr}{\mathrm}
\newcommand{\mbb}{\mathbb}

\newtheorem{theorem}{Theorem}

\definecolor{applegreen}{rgb}{0.55, 0.71, 0.0}

\newcommand{\cor}[1]{{\color{red}#1}}

\usepackage{algorithm}  
\usepackage[noend]{algpseudocode}  
\newcommand{\algorithmfootnote}[2][\footnotesize]{%
  \let\old@algocf@finish\@algocf@finish
  \def\@algocf@finish{\old@algocf@finish
    \leavevmode\rlap{\begin{minipage}{\linewidth}
    #1#2
    \end{minipage}}%
  }%
}
\usepackage{xparse}
\NewDocumentCommand{\LeftComment}{s m}{%
  \Statex \IfBooleanF{#1}{\hspace*{\ALG@thistlm}}\(\triangleright\) #2}
\algnewcommand{\LineComment}[1]{\Statex // #1}

\usepackage{footmisc}
\usepackage{cleveref}

\begin{document}
\title{Entanglement-enabled advantage for learning a bosonic random displacement channel}
\author{Changhun Oh}
\thanks{These authors contributed equally to this work: C.O. (\href{mailto:changhun0218@gmail.com}{changhun0218@gmail.com}); S.C. (\href{mailto:csenrui@uchicago.edu}{csenrui@uchicago.edu}).}
\affiliation{Pritzker School of Molecular Engineering, The University of Chicago, Chicago, Illinois 60637, USA}
\affiliation{Department of Physics, Korea Advanced Institute of Science and Technology, Daejeon 34141, Korea}
\author{Senrui Chen}
\thanks{These authors contributed equally to this work: C.O. (\href{mailto:changhun0218@gmail.com}{changhun0218@gmail.com}); S.C. (\href{mailto:csenrui@uchicago.edu}{csenrui@uchicago.edu}).}
\affiliation{Pritzker School of Molecular Engineering, The University of Chicago, Chicago, Illinois 60637, USA}
\author{Yat Wong}
\affiliation{Pritzker School of Molecular Engineering, The University of Chicago, Chicago, Illinois 60637, USA}
\author{Sisi Zhou}
\affiliation{Perimeter Institute for Theoretical Physics, Waterloo, Ontario N2L 2Y5, Canada}
\affiliation{Institute for Quantum Information and Matter, California Institute of Technology, Pasadena, CA 91125, USA}
\affiliation{Department of Physics and Astronomy and Institute for Quantum Computing, University of Waterloo, Ontario N2L 2Y5, Canada}
\author{Hsin-Yuan Huang}
\affiliation{Institute for Quantum Information and Matter,
California Institute of Technology, Pasadena, CA 91125, USA}
\affiliation{Center for Theoretical Physics,
Massachusetts Institute of Technology, Cambridge, MA 02139, USA}
\affiliation{Google Quantum AI, Venice, CA, USA}
\author{Jens A.H. Nielsen}
\author{Zheng-Hao Liu}
\author{Jonas S. Neergaard-Nielsen}
\author{Ulrik L. Andersen}
\affiliation{Center for Macroscopic Quantum States (bigQ), Department of Physics,
Technical University of Denmark, Building 307, Fysikvej, 2800 Kgs. Lyngby, Denmark}
\author{Liang Jiang}
\email{liang.jiang@uchicago.edu}
\affiliation{Pritzker School of Molecular Engineering, The University of Chicago, Chicago, Illinois 60637, USA}
\author{John Preskill}
\email{preskill@caltech.edu}
\affiliation{Institute for Quantum Information and Matter,
California Institute of Technology, Pasadena, CA 91125, USA}

\date{\today}
\begin{abstract}

We show that quantum entanglement can provide an exponential advantage in learning properties of a bosonic continuous-variable (CV) system. 
The task we consider is estimating a probabilistic mixture of displacement operators acting on $n$ bosonic modes, called a random displacement channel.  We prove that if the $n$ modes are not entangled with an ancillary quantum memory, then the channel must be sampled a number of times exponential in $n$ in order to estimate its characteristic function to reasonable precision; this lower bound on sample complexity applies even if the channel inputs and measurements performed on channel outputs are chosen adaptively. On the other hand, we present a simple entanglement-assisted scheme that only requires a number of samples independent of $n$, given a sufficient amount of squeezing. This establishes an exponential separation in sample complexity. We then analyze the effect of photon loss and show that the entanglement-assisted scheme is still significantly more efficient than any lossless entanglement-free scheme under mild experimental conditions. Our work illuminates the role of entanglement in learning continuous-variable systems and points toward experimentally feasible demonstrations of provable entanglement-enabled advantage using CV quantum platforms.
\end{abstract}
\maketitle

Quantum science and technology holds promise to revolutionize how we understand and interact with nature, enabling computational speedups~\cite{nielsen2002quantum}, classically impossible communication tasks~\cite{gisin2007quantum, kimble2008quantum}, and measurements with unprecedented sensitivity~\cite{giovannetti2006quantum, giovannetti2011advances, polino2020photonic}.
Rapid progress during the noisy intermediate-scale quantum (NISQ) era~\cite{preskill2018quantum} has brought these promises closer to reality, but the challenge remains to demonstrate rigorous quantum advantage for practical problems.

Over the past few years, there has been ongoing theoretical and experimental progress in exploring quantum computational advantage~\cite{aaronson2011computational, boixo2018characterizing, arute2019quantum, wu2021strong, zhong2020quantum, zhong2021phase, madsen2022quantum, morvan2023phase, deng2023gaussian}.
Another recent line of research seeks quantum advantage in learning~\cite{huang2021information, huang2022quantum, chen2022exponential,caro2022learning,bubeck2020entanglement,aharonov2022quantum,chen2021quantum,rossi2022quantum}, revealing that access to quantum memory enables us to learn properties of nature more efficiently.
Specifically, Refs.~\cite{huang2022quantum,chen2022exponential} establish a framework for proving exponential separation in sample complexity between learning with and without a coherently controllable quantum memory. 
In contrast to its computational counterpart, this entanglement-enabled advantage in learning can be proven without invoking computational assumptions and can sometimes be more experimentally accessible. A proof-of-principle experiment has been conducted on Google's superconducting quantum processor using $40$ qubits~\cite{huang2022quantum}. 

Most learning tasks studied so far are restricted to discrete-variable~(DV) systems. 
It is natural to ask whether entanglement-enabled advantage can also be realized for learning properties of bosonic continuous-variable (CV) systems.
This is particularly important because 
CV systems are ubiquitous in nature and have many applications in quantum information science, such as quantum sensing~\cite{braunstein2005quantum, weedbrook2012gaussian, serafini2017quantum, polino2020photonic,aaronson2011computational}.
However, generalizing the results in DV systems to CV systems is challenging because bosonic systems have infinite-dimensional Hilbert spaces, complicating rigorous complexity analysis of learning.
Recent progress has been achieved in studies of entanglement-enhanced learning of CV-state characteristic functions~\cite{wu2023quantum};
however, the lower bounds obtained so far apply to a restricted class of learning strategies rather than to general entanglement-free schemes. There have also been recent works studying applications of statistical learning theory to CV systems~\cite{becker2024classical,gandhari2024precision,rosati2022learning}.

\begin{figure}[t]
\includegraphics[width=240px]{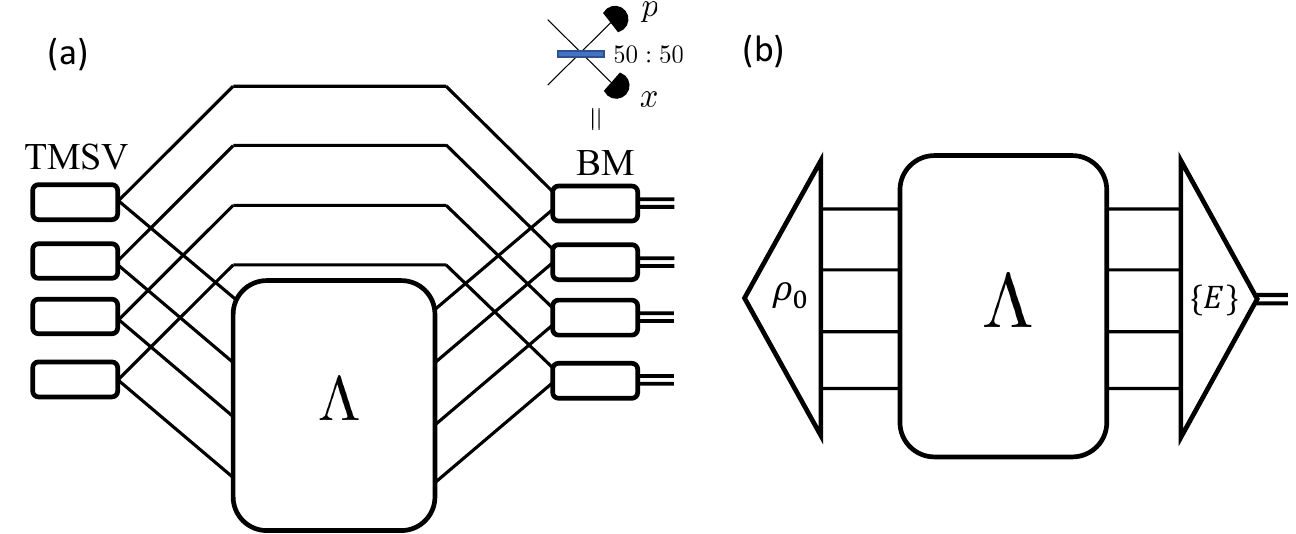}
\caption{Schemes for learning an $n$-mode random displacement channel $\Lambda$. (a) TMSV+BM, a specific entanglement-assisted~(EA) scheme. (b) General entanglement-free~(EF) scheme. See the main text for more details.}
\label{fig:scheme}
\end{figure}

Characterizing displacements applied to CV systems has many applications, including force sensing~\cite{gavartin2012hybrid}, dark matter search~\cite{sikivie1983experimental,brady2022entangled, backes2021quantum}, gravitational wave detection~\cite{ganapathy2023broadband}, and Raman scattering spectroscopy~\cite{de2020quantum}; therefore this task has drawn substantial recent attention~\cite{gardner2024stochastic, valahu2024benchmarking}.
In this work, we rigorously establish an entanglement-enabled advantage in learning a probabilistic mixture of $n$-mode displacement operations, called a \textit{random displacement channel}.
%
Specifically, we prove that any schemes without ancillary quantum memory (\textit{i.e.}, entanglement-free) require exponentially many samples in $n$ to learn the channel's characteristic function with good precision and high success probability.
In contrast, we present a simple scheme utilizing entanglement with ancillary quantum memory~(\textit{i.e.}, entanglement-assisted) that can achieve the same learning task with sample complexity independent of $n$, given access to two-mode squeezed vacuum~(TMSV) states with sufficiently large squeezing parameter and Bell measurements~(BM).
This establishes an exponential separation between learning with and without entanglement in the bosonic system~(see Fig.~\ref{fig:scheme}). 


Notably, our entanglement-assisted scheme achieves an exponential advantage using only Gaussian input state preparation and measurements which are experimentally feasible.
Furthermore, we demonstrate the robustness of the entanglement-enabled advantage under realistic experimental conditions by analyzing the effects of photon loss, phase noise, and crosstalk, the most common noise sources in optical platforms, suggesting that for squeezing parameters and noise rates achievable in a state-of-the-art experiment, the advantage remains significant and thus can be realized shortly.

For DV systems, existing techniques for proving quantum advantages in learning typically involve characterizing the hardness of distinguishing a given state or channel from the maximally mixed state or completely depolarizing channel~\cite{huang2021information, huang2022quantum, chen2022exponential, chen2021quantum}. These states and channels are ill-defined for CV systems; therefore we needed a different approach to establish an entanglement-enabled advantage without making reference to unphysical states
(see the proof of Theorem~\ref{th:lower}).
Another distinctive feature of our results is that our entanglement-free lower bound (Theorem~\ref{th:lower}) holds even for schemes that use input states with arbitrarily high energy (\emph{i.e.}, occupying an arbitrarily large Hilbert space), while the entanglement-assisted upper bound (Theorem~\ref{th:upper}) only uses finite energy input states.


\medskip
\textit{Problem Setup.{\textemdash}}
We consider the task of learning an $n$-mode random displacement channel characterized by a probability distribution $p(\alpha)$ with $\alpha \in\mathbb{C}^n$, which transforms an input state $\hat{\rho}$ as
\begin{align}\label{eq:rdc}
    \Lambda(\hat{\rho})=\int d^{2n}\alpha ~p(\alpha) \hat{D}(\alpha)\hat{\rho} \hat{D}^\dagger(\alpha),
\end{align}
where $\hat{D}(\alpha)\coleq\otimes_{i=1}^n \hat{D}(\alpha_i)$ and $\hat{D}(\alpha_i)\coleq\exp(\alpha_i\hat{a}_i^\dagger-\alpha_i^*\hat{a}_i)$ is the displacement operator for the $i$th mode.
It can be equivalently described by the characteristic function of $p(\alpha)$, \emph{i.e.}, its Fourier transform, as (see SM~S1~\cite{supple} for the derivation) 
\begin{align}
    &\Lambda(\hat{\rho})
    =\frac{1}{\pi^n} \int d^{2n}\beta ~\lambda(\beta) \Tr[\hat{\rho} \hat{D}(\beta)] \hat{D}^\dagger(\beta), \\ 
    &\lambda(\beta)\coleq\int d^{2n}\alpha ~p(\alpha)e^{\alpha^\dagger\beta-\beta^\dagger\alpha}.
\end{align}
Here, because of the Fourier relation, $\lambda(\beta)$ with a large $\beta$ contributes to rapidly oscillating $p(\alpha)$.
Since the domain of $\beta$ is infinite, we will focus on a restricted finite domain specified later.
The goal is to learn the channel by estimating the characteristic function $\lambda(\beta)$;
this is distinct from identifying a particular displacement drawn from the distribution $p(\alpha)$.
The value of $\beta$ of $\lambda(\beta)$ to be estimated is revealed only after all measurements are completed.


We focus on the separation between two types of learning schemes for the random displacement channel distinguished by whether the scheme uses entanglement between the system and an ancilla~(see Fig.~\ref{fig:scheme}). 
Throughout this work, we define an entanglement-free scheme as being both ancilla-free and concatenation-free, {\it i.e.}, the channel's output is measured destructively after each channel use. 
However, it is allowed to be adaptive; for each channel use, the input to the channel and the measurement may depend on measurement outcomes obtained in earlier rounds. 
This scenario is similar to Refs.~\cite{huang2021information,aharonov2022quantum}. 
Several recent works have obtained lower bounds on learning DV channels that hold even with concatenation~\cite{chen2021quantum,chen2023futility,chen2023tight}, but for simplicity, we focus on concatenation-free schemes.

\medskip
\textit{Schemes.{\textemdash}}
Now, we present an entanglement-assisted scheme~(see Fig.~\ref{fig:scheme}) inspired by Pauli channel learning proposed in Ref.~\cite{chen2021quantum}.
Consider an $n$-mode random displacement channel $\Lambda_{B}$ acting on the $n$-mode system $B$.
To learn this channel, we prepare $n$ CV Bell states with a finite squeezing parameter $r$, {\it i.e.} TMSV states, and half of the states go through the channel while the other half stays in quantum memory.
Finally, we measure the output state by CV BM, implemented by passing through 50:50 beam splitters and performing homodyne measurement on output ports along different quadratures~\cite{serafini2017quantum}.
Formally, the BM POVM element labeled by $\{\zeta \in \mathbb{C}^n\}$ has the following form: $(I\otimes\hat{D}(\zeta))|\Psi\rangle\langle \Psi|(I\otimes \hat{D}^\dagger(\zeta))/\pi^n$; here $|\Psi\rangle$ denotes the tensor product of $n$ infinitely squeezed TMSV states, each proportional to $\sum_{k=0}^\infty |k\rangle|k\rangle$ when expressed in the Fock basis.
To see how to learn a random displacement channel using this TMSV+BM scheme, we invoke the probability of obtaining outcome $\zeta$ from BM~(see SM~S2~A~\cite{supple} for the derivation):
\begin{align}\label{eq:pea}
    p_{EA}(\zeta)&=\frac{1}{\pi^{2n}}\int d^{2n}\alpha~\lambda(\alpha)e^{-e^{-2r}|\alpha|^2}e^{\alpha^\dagger\zeta-\zeta^\dagger\alpha}.
\end{align}
Fourier transforming to invert this relation, we obtain
\begin{equation}
\begin{aligned}\label{eq:lambda-from-p-EA}
    \lambda(\beta)
    &=e^{e^{-2r}|\beta|^2}\int d^{2n}\zeta~p_{EA}(\zeta)e^{\zeta^\dagger\beta-\beta^\dagger\zeta}\coleq e^{e^{-2r}|\beta|^2}\lambda_{EA}(\beta).
\end{aligned}
\end{equation}
This expression indicates that given $N$ outcomes $\{\zeta^{(i)}\}_{i=1}^N$ from the TMSV+BM scheme, $\tilde{\lambda}(\beta) \coleq \frac1Ne^{e^{-2r}|\beta|^2}\sum_{i=1}^N e^{\zeta^{(i)\dagger}\beta-\beta^\dagger\zeta^{(i)}}$ is an unbiased estimator of $\lambda(\beta)$.
Note that the same set of samples can be used to estimate 
$\lambda(\beta)$ for different values of $\beta$ just by modifying the estimator.
Using the Hoeffding's bound, we prove the following theorem (see SM~S2~A for the proof):
\begin{theorem}\label{th:upper}
    For any $n$-mode random displacement channel $\Lambda$, after the TMSV+BM scheme with squeezing parameter $r$ has learned from $N$ copies of $\Lambda$, and then received a query $\beta\in\mbb C^n$, it can provide an estimator $\tilde\lambda(\beta)$ of $\Lambda$'s characteristic function $\lambda(\beta)$ such that $|\tilde\lambda(\beta)-\lambda(\beta)|\le \epsilon$ with probability at least $1-\delta$, with the 
    number of samples $N = 8e^{2e^{-2r}|\beta|^2}\epsilon^{-2}\log4\delta^{-1}$.
\end{theorem}
In Theorem~\ref{th:lower} below we formulate a lower bound on the sample complexity for general entanglement-free schemes. First, though, 
let us compare the entanglement-assisted TMSV+BM scheme with a particular entanglement-free scheme that uses the vacuum state as input and heterodyne detection (Vacuum+Heterodyne).
Here, heterodyne detection is defined as a projection onto the (overcomplete) basis of coherent states, i.e., $|\zeta\rangle\langle \zeta|/\pi^n$ with $\zeta\in\mathbb{C}^n$.
This specific scheme helps us to understand the limitations of more general entanglement-free schemes which are captured more fully by Theorem~\ref{th:lower}.

In this scheme, the probability of obtaining the outcome $\zeta$ is (see SM~S2~B~\cite{supple})
\begin{align}\label{eq:pef}
    p_{{VH}}(\zeta)
    =\frac{1}{\pi^{2n}}\int d^{2n}\alpha~ \lambda(\alpha)e^{\alpha^\dagger\zeta-\zeta^\dagger\alpha}e^{-|\alpha|^2}.
\end{align}
In fact, the Vacuum+Heterodyne scheme can be understood as the TMSV+BM scheme with $r=0$.
Inverting this relation by Fourier transforming, we again express the channel's characteristic function by the measurement probability distribution:
\begin{align}\label{eq:lambda-from-p-EF}
    \lambda(\beta)
    =e^{|\beta|^2}\int d^{2n}\zeta~ p_{{VH}}(\zeta)e^{\zeta^\dagger\beta-\beta^\dagger\zeta}
    \coleq e^{|\beta|^2}\lambda_{VH}(\beta),
\end{align}
which yields another unbiased estimator $\tilde{\lambda}(\beta)\coleq\frac{1}{N}e^{|\beta|^2}\sum_{i=1}^N e^{\zeta^{(i)\dagger}\beta-\beta^\dagger\zeta^{(i)}}$ given $N$ outcomes $\{\zeta^{(i)}\}_{i=1}^N$. Comparing to \eqref{eq:lambda-from-p-EA}, the $r$-dependent prefactor is missing from \eqref{eq:lambda-from-p-EF}.
Specifically, if we confine $\beta$ to $|\beta|^2\leq \kappa n$ with a constant $\kappa>0$, we obtain upper bounds on the sample complexity for achieving an error $\epsilon$ with success probability $1{-}\delta$ using each scheme:
\begin{align}
    N_{EA}=O(e^{2e^{-2r}\kappa n}\epsilon^{-2}\log \delta^{-1})\label{eq:ideal_upper_bound}, 
    N_{VH}=O(e^{2\kappa n}\epsilon^{-2}\log \delta^{-1}).
\end{align}
In particular, if we choose $r=\Omega(\log n)$, the sample complexity of the entanglement-assisted scheme becomes independent of the number of modes $n$, while our upper bound on that of the entanglement-free scheme increases exponentially with $n$.
Since the accessible squeezing parameter is bounded in practice, though, we will consider the sample complexities for constant $r$ below.

\begin{figure}[t]
\includegraphics[width=200px]{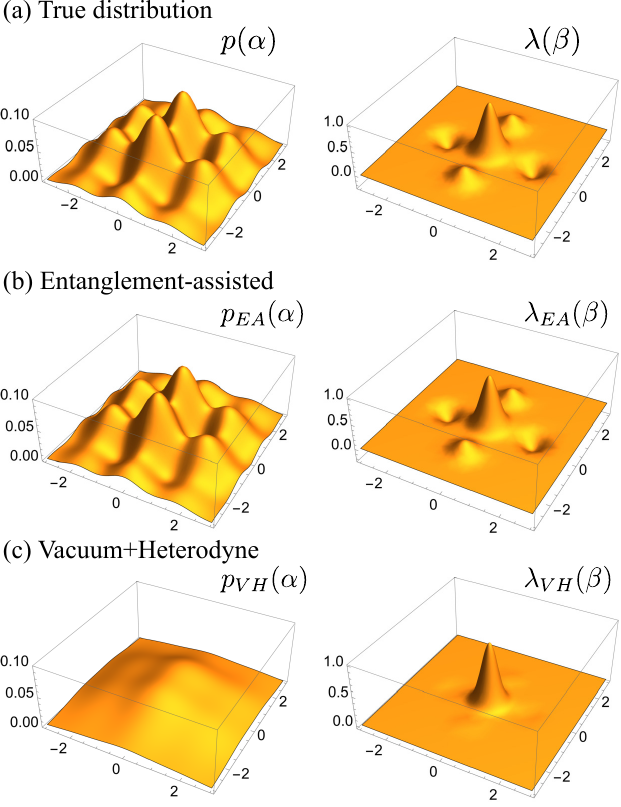}
\caption{Comparison between (a) the true distribution, (b) TMSV+BM, and (c) Vacuum+Heterodyne strategies.
The left panel represents the probability distribution of the true distribution and measurement probability distributions for each scheme.
The right panel represents the characteristic function of the probability distributions.}
\label{fig:comparison}
\end{figure}

To illustrate the difference, we compare the two strategies with a single-mode example, characterized by
\begin{align}
    p(\alpha)&=\frac{2\sigma^2}{\pi} e^{-2\sigma^2|\alpha|^2}[\cos^2(\alpha_r \gamma_i-\alpha_i \gamma_r) +\sin^2(\alpha_r \gamma_r+\alpha_i \gamma_i)], \\
    \lambda(\beta)&=e^{-\frac{|\beta|^2}{2\sigma^2}}+\frac{1}{4}e^{-\frac{|\beta-\gamma|^2}{2\sigma^2}}+\frac{1}{4}e^{-\frac{|\beta+\gamma|^2}{2\sigma^2}} \nonumber \\ 
    &~~~~~-\frac{1}{4}e^{-\frac{|\beta-i\gamma|^2}{2\sigma^2}}-\frac{1}{4}e^{-\frac{|\beta+i\gamma|^2}{2\sigma^2}},
\end{align}
with $\sigma=0.3$, $\gamma_r=1.6$, $\gamma_i=0$ ($\gamma\coleq\gamma_r+i\gamma_i$), and $r=2$ for the TMSV+BM scheme.
Figure~\ref{fig:comparison}, presenting the underlying output probability distributions and their characteristic functions from Eqs.~\eqref{eq:pea},\eqref{eq:lambda-from-p-EA},\eqref{eq:pef}, and \eqref{eq:lambda-from-p-EF}, clearly shows that in the TMSV+BM scheme with a sufficient squeezing parameter, the resultant probability distribution and characteristic function are almost identical to the ideal case.
However, for the Vacuum+Heterodyne scheme, the vacuum noise distorts the initial probability distribution so significantly that we cannot see the signal clearly, making it harder to estimate the original characteristic function.

\medskip
\textit{Lower bound.{\textemdash}}
Our upper bound on the sample complexity of the Vacuum+Heterodyne scheme scales exponentially with $n$. Can this scaling be improved using 
more advanced entanglement-free schemes, such as general-dyne detection or photon-number resolving measurements~\cite{schuster2007resolving}, or by non-Gaussian resources like GKP states~\cite{gottesman2001encoding}?
Here, using information-theoretic methods, we prove an exponential sample complexity lower bound for any entanglement-free scheme.
This highlights the indispensable role of entanglement for efficiently learning random displacement channels.
Our result is as follows:
\begin{theorem}\label{th:lower}
       Let $\Lambda$ be an arbitrary $n$-mode random displacement channel ($n\ge 8$) and consider an entanglement-free scheme that uses $N$ copies of $\Lambda$. After all measurements are completed, the scheme receives the query $\beta\in \mathbb{C}^n$ and returns an estimate $\tilde\lambda(\beta)$ of $\Lambda$'s characteristic function $\lambda(\beta)$. Suppose that, with success probability at least $2/3$, $|\tilde{\lambda}(\beta)-\lambda(\beta)|\le \epsilon\le 0.24$ for all $\beta$ such that $|\beta|^2 \le n\kappa$ and all $\Lambda$. Then $N\ge 0.01\epsilon^{-2}(1+1.98\kappa)^{n}$.
      
\end{theorem}
\noindent Here, the choice of success probability $2/3$ is arbitrary and can be easily amplified. Comparing with the entanglement-assisted sample complexity given in Eq.~\eqref{eq:ideal_upper_bound}, Theorem~\ref{th:lower} establishes a separation exponential in $n$ for cutoff coefficient $\kappa=O(1)$ and squeezing parameter $r=\Omega(\log n)$.
The intuition underlying this theorem is that displacement operators do not generally commute with each other. Consequently, entanglement-free measurements can resolve $\lambda(\beta)$ for only a small portion of $\beta$ space.
We sketch the proof below and leave the full details to SM~S3~\cite{supple}.

\begin{proof}[Proof Sketch]
We begin by defining the following family of ``$3$-peak'' random displacement channels $\bm\Lambda^{\epsilon,\sigma}_\mr{3\mhyphen peak}=\{\Lambda_{\gamma}\}_{\gamma\in\mathbb{C}^n}$ whose characteristic functions and distributions of displacement are, respectively,
\begin{align}
    \lambda_{\gamma}(\beta)
    &=
    e^{-\frac{|\beta|^2}{2\sigma^2}}+2i\epsilon_0 e^{-\frac{|\beta-\gamma|^2}{2\sigma^2}}-2i\epsilon_0 e^{-\frac{|\beta+\gamma|^2}{2\sigma^2}},\\
    p_\gamma(\alpha) &\propto e^{-2\sigma^2|\alpha|^2}\left(1+4\epsilon_0\sin(2(\gamma_i\alpha_r-\gamma_r\alpha_i))\right).
\end{align}
\noindent with positive parameters $\sigma$ and $\epsilon\coleq0.98\epsilon_0$. 
Note that the channel with $\sigma\to 0$ and $\gamma=0$ resembles the completely depolarizing channel used in Refs.~\cite{chen2021quantum,huang2022quantum}, which outputs the maximally mixed state in the infinite-dimensional Hilbert space, $\propto \sum_{k=0}^\infty |k\rangle\langle k|$, that is mathematically ill-defined and thus unphysical; however, because $\sigma$ is strictly larger than zero throughout the proof, the channels are well-defined.
We will show that, even with the prior knowledge that the channel is from this family, it is still hard for entanglement-free schemes to achieve the learning tasks.

The key idea is to reduce learning to hypothesis testing. Consider the following game between Alice and Bob: Alice chooses one of two hypotheses with equal probability: (1)~Set $\Lambda=\Lambda_0$; (2)~Set $\Lambda=\Lambda_\gamma$ for $\gamma$ sampled from a Gaussian distribution whose variance is determined by $\kappa$.
Next, Alice allows Bob to use the channel $\Lambda$ $N$ times, and Bob uses any entanglement-free schemes to learn from these channel uses. 
After Bob has finished all quantum measurements and keeps only classical data, Alice reveals some auxiliary information to Bob, and Bob then decides whether Alice has chosen (1) or (2).

Using a learning scheme satisfying the assumptions of Theorem~\ref{th:lower}, Bob can guess correctly with high probability;
Thus, the outcome distributions of Bob's scheme under hypotheses (1) and (2) must have a large total variation distance~(TVD). Meanwhile, we prove that the contribution from each use of $\Lambda$ to the TVD is exponentially small using a new technique inspired by Ref.~\cite{caves2004fidelity}, which derived the maximum fidelity of Gaussian random displacement channels.
Therefore, the number of channel uses $N$ must be exponentially large to ensure a large TVD, which gives us the desired lower bound on the sample complexity.
\end{proof}

\begin{figure*}[t]
	\centering
        \includegraphics[width=0.95\linewidth]{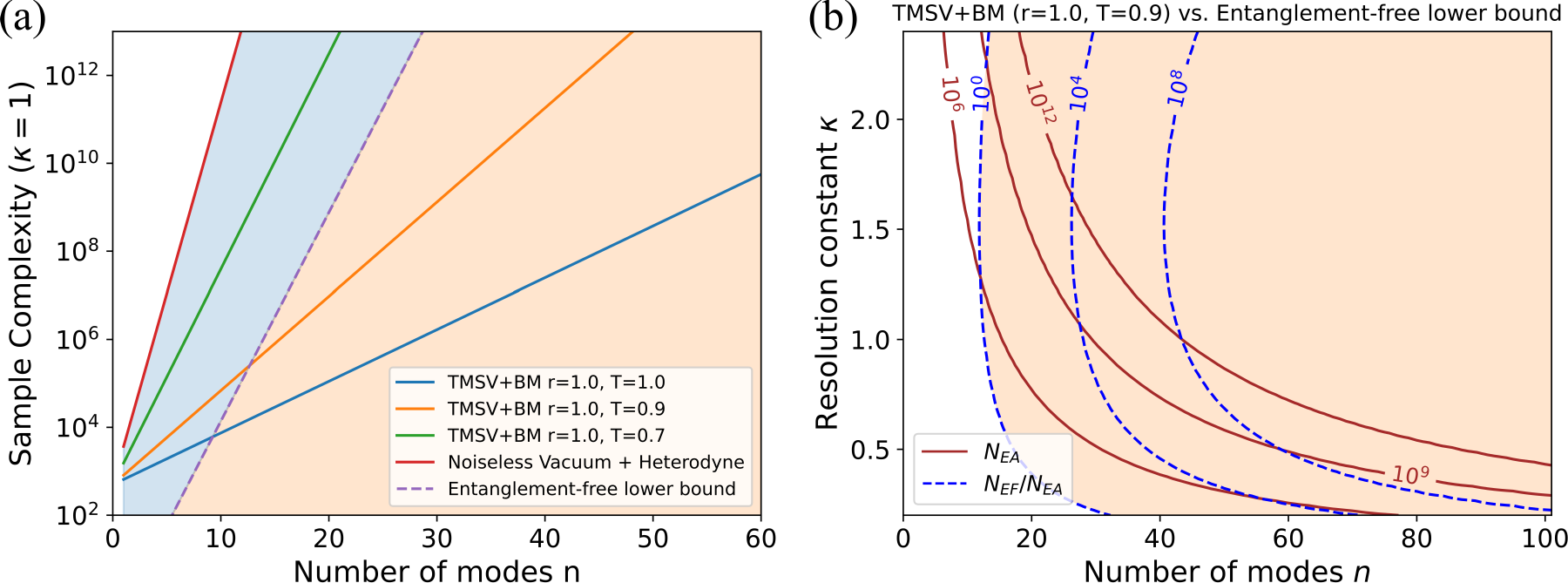}
        \caption{
        (a)~Comparison of TMSV+BM (with different loss rates), Vacuum+Heterodyne, and the entanglement-free lower bound at $\kappa=1$. The task is to estimate any $\lambda(\beta)$ such that $|\beta|^2\le\kappa n$ with precision $\varepsilon=0.2$ and success probability $1-\delta=2/3$. The orange region represents a rigorous advantage over all entanglement-free schemes. The blue region represents an advantage over noiseless Vacuum+Heterodyne.
        (b)~Comparison of the TMSV+BM scheme with squeezing parameter $r=1.0$ and loss rate $1-T=0.1$ with the entanglement-free lower bound of Theorem~\ref{th:lower}. (See SM~S3~A for further practical considerations.) The task is the same as (a). 
        The brown solid contour lines represent the sample complexity of TMSV+BM given by Theorem~\ref{th:lossy}. The blue dashed contour lines represent the ratio of sample complexity between the entanglement-free lower bound and TMSV+BM, indicating the entanglement-enabled advantage.
	}
	\label{fig:result}
\end{figure*}

\medskip

\textit{Effect of noise.{\textemdash}}
Now, for practical applications, we analyze the effect of dominant noise sources in optical platforms on the entanglement-assisted scheme and demonstrate its robustness against them, implying that the advantage is realizable shortly.
We discuss the photon-loss effect in the main text, while phase diffusion and measurement crosstalk are covered in SM~S2~D and E.
Photon loss transforms the relevant bosonic operator $\hat{a}$ to $\sqrt{T}\hat{a}+\sqrt{1-T}\hat{e}$, where $T$ is the transmission rate and $\hat{e}$ is the environmental mode, \textit{i.e.}, $1-T$ is the loss rate.
We consider two different places where the loss occurs: one is before applying the channel with loss rate $1-T_b$ to model the preparation imperfection, and the other is after applying the channel and before the perfect BM with loss rate $1-T_a$, which models an imperfect BM~\cite{serafini2017quantum}. 
As before, we derive the relation between the measurement probability distribution and the characteristic function of the channel (with appropriate rescaling of the phase):
\begin{align}
    \lambda(\beta)
    &= e^{e^{-2r_\mr{eff}}|\beta|^2}\int d^{2n}\zeta ~p_{loss}(\zeta) e^{(\zeta^\dagger\beta-\beta^\dagger\zeta)/\sqrt{T_a}},
\end{align}
where we define an effective squeezing parameter \begin{equation}\label{eq:effective_squeezing_parameter}
    r_\text{eff}\coleq -\frac{1}{2}\log\left(T_be^{-2r}+(1-T_b)+\frac{1-T_a}{T_a}\right),
\end{equation}
which incorporates the loss rates.
Because loss degrades the advantage from squeezing, the upper bound on sample complexity in Theorem~\ref{th:upper} is modified as
(see SM~S2~C for the proof):
\begin{theorem}\label{th:lossy}
    For the same task as in Theorem~\ref{th:upper},    
    a 
    TMVS+BM scheme with squeezing parameter $r$ and transmission rates before and after the channel $T_b$ and $T_a$ respectively, can estimate 
    any $\lambda(\beta)$ to error $\epsilon$ with success probability $1{-}\delta$ using the 
    number of samples $N=8e^{2e^{-2r_\text{eff}}|\beta|^2}\epsilon^{-2}\log4\delta^{-1}$, where $r_\mr{eff}$ is defined according to Eq.~\eqref{eq:effective_squeezing_parameter}.
\end{theorem}
\noindent 
Thus, when $|\beta|^2\leq \kappa n$ with a constant $\kappa>0$, $T_b=1-O(1/n)$, $T_a=1-O(1/n)$ and $r=\Omega(\log n)$, the sample complexity becomes $N=O(\epsilon^{-2}\log\delta^{-1})$ as in the lossless case.
For practically relevant squeezing and including loss 
prior to BM, we compare the sample complexity for the lossy TMSV+BM protocol and the lossless entanglement-free lower bound in Fig.~\ref{fig:result}, 
finding a significant entanglement-enabled advantage in realistic experimental settings. 
Specifically, for reasonable parameter choices such as squeezing parameter $r=1$, loss rate $10\%$, and $\kappa=O(1)$, we can achieve a factor of $10^4~(10^8)$ advantage for around $n=30~(60)$ modes.
Although the $10^9$ number of samples required to achieve the advantage seems large, the state-of-the-art quantum optics experiments~({\it e.g.}, Refs.~\cite{guo2020distributed, inoue2023toward}) can attain such number of samples in a reasonable time with high sampling rate up to 160~GHz.



\medskip
\textit{Discussion.{\textemdash}}
We proved that schemes exploiting entanglement with an ancillary quantum memory can learn $n$-mode random displacement channels with exponentially fewer samples compared to entanglement-free schemes. 
Our proof technique generalizes the information-theoretic framework for learning studied in DV quantum systems~\cite{huang2021information,chen2022exponential} to the CV setting.
We anticipate that many other results concerning DV systems can be generalized using similar methods to CV learning tasks, such as learning CV quantum states or other CV channels.
Also, our analysis suggests that the separation in sample complexity between entanglement-assisted and entanglement-free protocols may soon be realized experimentally.


Besides their theoretical interest, random displacement channels can be practically relevant in, \textit{e.g.}, modeling noise in bosonic systems. As in qubit cases~\cite{wallman2016noise}, we expect that noise tailoring methods can transform more general noise models into random displacement channels; therefore efficiently learning random displacement channels can be useful for benchmarking CV quantum systems~\cite{wu2019efficient,bai2018test,valahu2024benchmarking} and error mitigation.


Displacement estimation is also studied in quantum metrology~(see \textit{e.g.}~\cite{shi2023ultimate,zhuang2018distributed,xia2020demonstration}). A task often considered in metrology is learning an unknown \emph{unitary} displacement or phase transformation acting independently on each mode \cite{duivenvoorden2017single, zhuang2018distributed, guo2020distributed, oh2020optimal, oh2022distributed, kwon2022quantum} whereas the task analyzed in this work is learning an unknown \emph{mixture} of multimode displacements. 
Furthermore, while the goal in metrology is typically to learn one or a few parameters, in our case, the parameter space is very large. 
Therefore, the methodology in the two settings is quite different. 
Connections between metrology and bosonic channel learning are worthy of further exploration.

\bigskip

\begin{acknowledgments}
   We thank Mankei Tsang, Yuxin Wang, Ronald de Wolf, Mingxing Yao, Ming Yuan for insightful discussions.
   C.O., S.C., Y.W., L.J. acknowledge support from the ARO(W911NF-23-1-0077), ARO MURI (W911NF-21-1-0325), AFOSR MURI (FA9550-19-1-0399, FA9550-21-1-0209), NSF (OMA-1936118, ERC-1941583, OMA-2137642), NTT Research, Packard Foundation (2020-71479). J.P. acknowledges support from the U.S. Department of Energy Office of Science, Office of Advanced Scientific Computing Research (DE-NA0003525, DE-SC0020290), the U.S. Department of Energy, Office of Science, National Quantum Information Science Research Centers, Quantum Systems Accelerator, and the National Science Foundation (PHY-1733907). The Institute for Quantum Information and Matter is an NSF Physics Frontiers Center. S.Z. acknowledges funding provided by the Institute for Quantum Information and Matter and Perimeter Institute for Theoretical Physics, a research institute supported in part by the Government of Canada through the Department of Innovation, Science and Economic Development Canada and by the Province of Ontario through the Ministry of Colleges and Universities. J.A.H.N, Z.-H.L., J.S.N-N. and U.L.A acknowledge support from DNRF (bigQ, DNRF142), IFD (PhotoQ, 1063-00046A) and EU (CLUSTEC, ClusterQ ERC-101055224, GTGBS MC-101106833).
   This research was supported by Quantum Technology R\&D Leading Program~(Quantum Computing) (RS-2024-00431768) through the National Research Foundation of Korea~(NRF) funded by the Korean government (Ministry of Science and ICT~(MSIT))
\end{acknowledgments}


\bibliography{reference.bib}

\end{document}


\title{Entanglement-enabled advantage for learning a bosonic random displacement channel: Supplemental Material}

\author{Changhun Oh}
\thanks{These authors contributed equally to this work: C.O. (\href{mailto:changhun0218@gmail.com}{changhun0218@gmail.com}); S.C. (\href{mailto:csenrui@uchicago.edu}{csenrui@uchicago.edu}).}
\affiliation{Pritzker School of Molecular Engineering, The University of Chicago, Chicago, Illinois 60637, USA}
\affiliation{Department of Physics, Korea Advanced Institute of Science and Technology, Daejeon 34141, Korea}
\author{Senrui Chen}
\thanks{These authors contributed equally to this work: C.O. (\href{mailto:changhun0218@gmail.com}{changhun0218@gmail.com}); S.C. (\href{mailto:csenrui@uchicago.edu}{csenrui@uchicago.edu}).}
\affiliation{Pritzker School of Molecular Engineering, The University of Chicago, Chicago, Illinois 60637, USA}
\author{Yat Wong}
\affiliation{Pritzker School of Molecular Engineering, The University of Chicago, Chicago, Illinois 60637, USA}
\author{Sisi Zhou}
\affiliation{Perimeter Institute for Theoretical Physics, Waterloo, Ontario N2L 2Y5, Canada}
\affiliation{Institute for Quantum Information and Matter, California Institute of Technology, Pasadena, CA 91125, USA}
\affiliation{Department of Physics and Astronomy and Institute for Quantum Computing, University of Waterloo, Ontario N2L 2Y5, Canada}
\author{Hsin-Yuan Huang}
\affiliation{Institute for Quantum Information and Matter,
California Institute of Technology, Pasadena, CA 91125, USA}
\affiliation{Center for Theoretical Physics,
Massachusetts Institute of Technology, Cambridge, MA 02139, USA}
\affiliation{Google Quantum AI, Venice, CA, USA}
\author{Jens A.H. Nielsen}
\author{Zheng-Hao Liu}
\author{Jonas S. Neergaard-Nielsen}
\author{Ulrik L. Andersen}
\affiliation{Center for Macroscopic Quantum States (bigQ), Department of Physics,
Technical University of Denmark, Building 307, Fysikvej, 2800 Kgs. Lyngby, Denmark}
\author{Liang Jiang}
\email{liang.jiang@uchicago.edu}
\affiliation{Pritzker School of Molecular Engineering, The University of Chicago, Chicago, Illinois 60637, USA}
\author{John Preskill}
\email{preskill@caltech.edu}
\affiliation{Institute for Quantum Information and Matter,
California Institute of Technology, Pasadena, CA 91125, USA}

\date{\today}

\maketitle
\tableofcontents

\section{Preliminary}
In this section, we provide some identities that are frequently used in Supplemental Material (more details can be found in Refs.~\cite{ferraro2005gaussian, weedbrook2012gaussian, serafini2017quantum}).
First, an elementary operator in $n$-mode bosonic system is $n$-mode displacement operator $\hat{D}(\beta)\coleq e^{\beta \hat{a}^\dagger-\beta^\dagger \hat{a}}$, where $\beta\coleq (\beta_1,\dots,\beta_n)^\text{T}\in\mathbb{C}^n$, $\hat{a}\coleq(\hat{a}_1,\dots,\hat{a}_n)^\text{T}$ and $\hat{a}^\dagger\coleq (\hat{a}_1^\dagger,\dots,\hat{a}_n^\dagger)^\text{T}$ are annihilation and creation operator of bosons, which follow the commutation relation $[\hat{a}_i,\hat{a}_j^\dagger]=\delta_{ij}$.
Displacement operator $\hat{D}(\beta)$ forms an orthogonal basis in the operator space; thus, any operator $\hat{O}$ can be expanded by displacement operators as
\begin{align}\label{eq:displacement_fourier}
    \hat{O}=\frac{1}{\pi^n}\int d^{2n}\beta \Tr[\hat{O}\hat{D}(\beta)]\hat{D}^\dagger(\beta),
\end{align}
where $\text{Tr}[\hat{O}\hat{D}(\beta)]$ is called the characteristic function of an operator $\hat{O}$.
The $n$-mode displacement operator has the following properties
\begin{align}
    \hat{D}^\dagger(\beta)=\hat{D}(-\beta),~~~\hat{D}^*(\beta)=\hat{D}(\beta^*),~~~\hat{D}^\text{T}(\beta)=\hat{D}(-\beta^*),~~~\Tr[\hat{D}(\beta)]=\pi^n \delta^{(2n)}(\beta),  \label{eq:dis1}
    \\ 
    \hat{D}(\beta_1)\hat{D}(\beta_2)=\hat{D}(\beta_1+\beta_2)e^{(\beta_2^\dagger \beta_1-\beta_1^\dagger\beta_2)/2},~~~\hat{D}(\alpha)\hat{D}^\dagger(\beta)\hat{D}^\dagger(\alpha)=\hat{D}^\dagger(\beta)e^{\alpha^\dagger\beta-\beta^\dagger\alpha}, \label{eq:dis2} \\
    \int\frac{d^{2n}\beta}{\pi^n}\hat{D}(\beta)O\hat{D}^\dagger(\beta)=\Tr[O]\mathds{1}
    \label{eq:dis3},
\end{align}
where the last identity is the twirling identity.
We also frequently use the following identity:
\begin{align}\label{eq:delta}
    \delta^{(2n)}(\alpha)&=\frac{1}{\pi^{2n}}\int d^{2n}\beta e^{\beta^\dagger\alpha-\alpha^\dagger\beta}.
\end{align}

Also, we employ the Wigner function of an operator $\hat{O}$ defined as
\begin{align}
    W_{O}(\alpha)
    =\frac{1}{\pi^{2n}}\int d^{2n}\beta \Tr[\hat{O} \hat{D}(\beta)]e^{\beta^\dagger\alpha-\alpha^\dagger \beta}.
\end{align}

In Sec.~\ref{app:fourier}, we show that for a given random displacement channel, which is defined by the probability distribution $p(\alpha)$, we can rewrite it by the Fourier transformation $\lambda(\beta)$ of $p(\alpha)$, i.e., its characteristic function, as
\begin{align}
    \Lambda(\hat{\rho})
    \coleq \int d^{2n}\alpha p(\alpha) \hat{D}(\alpha)\hat{\rho} \hat{D}^\dagger(\alpha)
    =\frac{1}{\pi^n} \int d^{2n}\beta \lambda(\beta) \Tr[\hat{\rho} \hat{D}(\beta)] \hat{D}^\dagger(\beta),
\end{align}
where the probability $p(\alpha)$ and the characteristic function $\lambda(\beta)$ follow the relation
\begin{align}
    \lambda(\beta) = \int d^{2n}\alpha~ p(\alpha) e^{\alpha^\dagger\beta-\beta^\dagger \alpha},~~
    p(\alpha) = \frac{1}{\pi^{2n}}\int d^{2n}\beta~ \lambda(\beta) e^{\beta^\dagger \alpha-\alpha^\dagger\beta}.
\end{align}

\subsection{Fourier relation}\label{app:fourier}
Here, we derive the expression of a random displacement channel characterized by a probability distribution $p(\alpha)$ by its characteristic function $\lambda(\beta)$.
To see the relation, we use the identity~\eqref{eq:displacement_fourier}.
Applying this for the density operator $\hat{\rho}$, we can show that
\begin{align}
    \Lambda(\hat{\rho})
    &=\int d^{2n}\alpha p(\alpha) \hat{D}(\alpha)\hat{\rho} \hat{D}^\dagger(\alpha)
    =\int d^{2n}\alpha p(\alpha) \hat{D}(\alpha)\left[\frac{1}{\pi^n}\int d^{2n}\beta \Tr[\hat{\rho} \hat{D}(\beta)]\hat{D}^\dagger(\beta)\right] \hat{D}^\dagger(\alpha) \\ 
    &=\frac{1}{\pi^n}\int d^{2n}\alpha \int d^{2n}\beta p(\alpha) \Tr[\hat{\rho} \hat{D}(\beta)] \hat{D}(\alpha)\hat{D}^\dagger(\beta) \hat{D}^\dagger(\alpha)\\
    &=\frac{1}{\pi^n}\int d^{2n}\alpha \int d^{2n}\beta p(\alpha) \Tr[\hat{\rho} \hat{D}(\beta)] \hat{D}^\dagger(\beta) e^{\alpha^\dagger\beta-\beta^\dagger\alpha} \\ 
    &=\frac{1}{\pi^n} \int d^{2n}\beta \lambda(\beta) \Tr[\hat{\rho} \hat{D}(\beta)] \hat{D}^\dagger(\beta),
\end{align}
where we used the identities from Eqs.~\eqref{eq:dis2},\eqref{eq:delta}.
Here, the last line renders the identity
\begin{align}
    \lambda(\beta)=\int d^{2n}\alpha p(\alpha)e^{\alpha^\dagger\beta-\beta^\dagger\alpha}.
\end{align}
Thus, $\lambda(\beta)$ is the Fourier transform of $p(\alpha)$.
Its inverse Fourier transformation gives
\begin{align}
    p(\alpha)=\frac{1}{\pi^{2n}}\int d^{2n}\beta \lambda(\beta)e^{\beta^\dagger \alpha-\alpha^\dagger\beta}.
\end{align}

\section{Derivation of output probability distributions of each scheme}

\subsection{Entanglement-assisted (TMSV+BM) schemes}\label{app:bell_finite}

In this section, we consider the two-mode squeezed vacuum and Bell measurement (TMSV+BM) strategy (see Fig.~1(a) in the main text.) and prove Theorem 1 in the main text by deriving the sample complexity of the strategy.
We assume a lossless system and a perfect Bell measurement, whereas the input squeezed state has a finite squeezing parameter $r$ since the input squeezing parameter $r$ is typically upper-bounded by a constant in practice.
We then analyze the effect of the imperfect measurement and loss in Sec.~\ref{app:combine}.

We now derive the probability of outcomes of the strategy, i.e., the outcomes obtained by applying an $n$-mode channel $\Lambda$ onto a product state of a subsystem of $n$ TMSV states $|\tilde{\Psi}\rangle$ and measuring in the Bell basis $|\Psi(\zeta)\rangle\langle \Psi(\zeta)|/\pi^n$ with $\zeta\in\mathbb{C}^n$:
\begin{align}
    p_{EA}(\zeta)&=\frac{1}{\pi^{n}}\Tr[(|\Psi(\zeta)\rangle\langle \Psi(\zeta)|_{AB})(\mathcal{I}_A\otimes \Lambda_{B})(|\tilde{\Psi}\rangle\langle \tilde{\Psi}|_{AB})].
\end{align}
To simplify the expression, we rewrite a TMSV state.
To do that, let us first consider a single-mode squeezed state $|r\rangle\coleq \hat{S}(r)|0\rangle$:
\begin{align}
    |r\rangle\langle r|
    =\frac{1}{\pi}\int d^2\alpha \hat{D}^\dagger(\alpha)\Tr[\hat{D}(\alpha)|r\rangle\langle r|]
    \coleq \frac{1}{\pi}\int d^2\alpha \hat{D}^\dagger(\alpha)f(\alpha,r),
\end{align}
where $\hat{S}(r)\coleq \exp[r(\hat{a}^{\dagger 2}-\hat{a}^2)/2]$ is the squeezing operation and we have defined
\begin{align}
    f(\alpha,r)
    &\coleq \Tr[\hat{D}(\alpha)|r\rangle\langle r|]
    =\langle 0|\hat{S}^\dagger(r) \hat{D}(\alpha)\hat{S}(r)|0\rangle
    =\langle 0|\hat{D}(\alpha \cosh r-\alpha^*\sinh r)|0\rangle \\ 
    &=\exp\left(-\frac{1}{2}|\alpha \cosh r-\alpha^*\sinh r|^2\right).
\end{align}
Here, we have used the relation, 
$\hat{S}^\dagger(r) \hat{a} \hat{S}(r)=\hat{a}\cosh r+\hat{a}^\dagger\sinh r.$
Using the fact that a TMSV state can be generated by injecting two single-mode squeezed states into the 50:50 beam splitter, i.e., $\hat{U}_{\text{BS}}(|r\rangle\langle r|\otimes |-r\rangle\langle -r|)\hat{U}_\text{BS}^\dagger=|\tilde{\Psi}(r)\rangle\langle \tilde{\Psi}(r)|$, where $\hat{U}_{\text{BS}}$ is 50:50 beam splitter, we can rewrite the TMSV state $|\tilde{\Psi}(r)\rangle$ as
\begin{align}
    |\tilde{\Psi}(r)\rangle\langle \tilde{\Psi}(r)|
    &=\frac{1}{\pi^2}\int d^2\alpha_1 d^2\alpha_2 \hat{U}_{\text{BS}}[\hat{D}^\dagger(\alpha_1)\otimes \hat{D}^\dagger(\alpha_2)]\hat{U}_{\text{BS}}^\dagger f(\alpha_1,r)f(\alpha_2,-r) \\ 
    &=\frac{1}{\pi^2}\int d^2\alpha_1 d^2\alpha_2\hat{D}^\dagger\left(\frac{\alpha_1-\alpha_2}{\sqrt{2}}\right)\otimes \hat{D}^\dagger\left(\frac{\alpha_1+\alpha_2}{\sqrt{2}}\right) f(\alpha_1,r)f(\alpha_2,-r) \\ 
    &=\frac{1}{\pi^2}\int d^2\omega_1 d^2\omega_2\hat{D}^\dagger(\omega_1)\otimes \hat{D}^\dagger(\omega_2) f\left(\frac{\omega_1+\omega_2}{\sqrt{2}},r\right)f\left(\frac{\omega_2-\omega_1}{\sqrt{2}},-r\right) \\
    &\coleq \frac{1}{\pi^2}\int d^2\omega_1 d^2\omega_2\hat{D}^\dagger(\omega_1)\otimes \hat{D}^\dagger(\omega_2) g(w_1,w_2,r),
\end{align}
where we defined
\begin{align}
    g(\omega_1,\omega_2,r)
    &\coleq  f\left(\frac{\omega_1+\omega_2}{\sqrt{2}},r\right)f\left(\frac{\omega_2-\omega_1}{\sqrt{2}},-r\right) \\ 
    &=\exp\left[-\frac{1}{2}\left[(|\omega_1|^2+|\omega_2|^2)\cosh2r-(\omega_1\omega_2+\omega_1^*\omega_2^*)\sinh{2r}\right]\right],
\end{align}
which is the characteristic function of the TMSV state $|\tilde{\Psi}(r)\rangle$ by Eq.~\eqref{eq:displacement_fourier}, i.e.,
\begin{align}
    g(\omega_1,\omega_2,r)\coleq \Tr[|\tilde{\Psi}(r)\rangle\langle\tilde{\Psi}(r)|(\hat{D}(\omega_1)\otimes \hat{D}(\omega_2))].
\end{align}
Multiple TMSV states are straightforward to generalize:
\begin{align}
    g(\omega_1,\omega_2,r)
    \coleq 
    \exp\left[-\frac{1}{2}\left[(|\omega_1|^2+|\omega_2|^2)\cosh2r-(\omega_1\cdot\omega_2+\omega_1^*\cdot\omega_2^*)\sinh{2r}\right]\right],
\end{align}
where $\omega_{1,2}$ are now $n$-dimensional complex vectors.
Especially when $\omega_2=\omega_1^*$, it reduces to
\begin{align}
    g(\omega_1,\omega_1^*,r)
    = 
    \exp\left(-e^{-2r}|\omega_1|^2\right).
\end{align}
Therefore, the input TMSV states can be written as
\begin{align}
    |\tilde{\Psi}(r)\rangle\langle \tilde{\Psi}(r)|
    =\frac{1}{\pi^{2n}}\int d^{2n}\omega_1 d^{2n}\omega_2\hat{D}^\dagger(\omega_1)\otimes \hat{D}^\dagger(\omega_2) g(w_1,w_2,r).
\end{align}

Similarly, by generalizing an entangled state with a single-mode ancilla to an entangled state with $n$ modes and infinite squeezing, the measurement POVM of CV Bell measurement can be written as~\cite{weedbrook2012gaussian, serafini2017quantum}
\begin{align}
    \frac{1}{\pi^n}|\Psi(\zeta)\rangle\langle \Psi(\zeta)|
    &=\frac{1}{\pi^n}(I\otimes \hat{D}(\zeta))|\Psi\rangle\langle \Psi|(I\otimes \hat{D}^\dagger(\zeta))
    =\int \frac{d^{2n}\omega}{\pi^{2n}}e^{\zeta^*\cdot \omega^*-\zeta\cdot\omega}\hat{D}^\dagger(\omega)\otimes \hat{D}^\text{T}(\omega),
\end{align}
where
\begin{align}
    |\Psi\rangle\langle \Psi|\coleq \int \frac{d^{2n}\omega}{\pi^{n}}\hat{D}^\dagger(\omega)\otimes \hat{D}^\text{T}(\omega)
\end{align}
corresponds to a multimode generalization of a TMSV state with infinite squeezing parameter, i.e., $|\Psi\rangle\propto\sum_{k=0}^\infty |k\rangle|k\rangle$.
Here, the normalization factor $1/\pi^n$ is introduced to ensure the completeness
\begin{align}
    \frac{1}{\pi^n}\int d^{2n}\zeta|\Psi(\zeta)\rangle\langle \Psi(\zeta)|=\mathds{1}\otimes \mathds{1}.
\end{align}

We now simplify the expression of the output probability distribution of the scheme:
\begin{align}\label{eq:bell_measure_finite}
    &p_{EA}(\zeta) \nonumber\\ 
    &=\frac{1}{\pi^n}\Tr\left[(|\Psi(\zeta)\rangle\langle \Psi(\zeta)|_{AB})(\mathcal{I}_A\otimes \Lambda_{B})(|\tilde{\Psi}\rangle\langle \tilde{\Psi}|_{AB})\right]\\
    &=\frac{1}{\pi^n}\Tr\bigg[\int\frac{d^{2n}\omega}{\pi^{n}}\frac{d^{2n}\beta_1d^{2n}\beta_2}{\pi^{2n}}g(\beta_1,\beta_2,r)e^{\zeta^*\cdot \omega^*-\zeta\cdot\omega}\hat{D}^\dagger(\omega)_A\otimes \hat{D}^\text{T}(\omega)_B(\mathcal{I}_A\otimes \Lambda_{B})(\hat{D}^\dagger(\beta_1)_A\otimes \hat{D}^\dagger(\beta_2)_B)\bigg].
\end{align}
Here, we have
\begin{align}
    (\mathcal{I}_A\otimes \Lambda_{B})(\hat{D}^\dagger(\beta_1)_A\otimes \hat{D}^\dagger(\beta_2)_B)
    &=\hat{D}^\dagger(\beta_1)_A\otimes \frac{1}{\pi^n}\int d^{2n}\omega\lambda(\omega)\Tr[\hat{D}^\dagger(\beta_2)_B \hat{D}(\omega)_{B}]\hat{D}^\dagger(\omega)_{B} \\ 
    &=\hat{D}^\dagger(\beta_1)_A\otimes \int d^{2n}\omega\lambda(\omega)\delta(\omega-\beta_2)\hat{D}^\dagger(\omega)_{B} \\ 
    &=\lambda(\beta_2) \hat{D}^\dagger(\beta_1)_A\otimes \hat{D}^\dagger(\beta_2)_{B}.
\end{align}
Thus, we can simplify Eq.~\eqref{eq:bell_measure_finite} as
\begin{align}
    &\frac{1}{\pi^n}\Tr\left[\int \frac{d^{2n}\omega}{\pi^{n}}\frac{d^{2n}\beta_1d^{2n}\beta_2}{\pi^{2n}}g(\beta_1,\beta_2,r)e^{\zeta^*\cdot \omega^*-\zeta\cdot\omega}\hat{D}^\dagger(\omega)_A\otimes \hat{D}^\text{T}(\omega)_B(\mathcal{I}_A\otimes \Lambda_{B})(\hat{D}^\dagger(\beta_1)_A\otimes \hat{D}^\dagger(\beta_2)_B)\right] \\
    &=\frac{1}{\pi^n}\int\frac{d^{2n}\omega}{\pi^{n}}\frac{d^{2n}\beta_1d^{2n}\beta_2}{\pi^{2n}}\lambda(\beta_2)g(\beta_1,\beta_2,r)e^{\zeta^*\cdot \omega^*-\zeta\cdot\omega}\Tr[\hat{D}^\dagger(\omega)_A\hat{D}^\dagger(\beta_1)_A]\Tr[\hat{D}^\text{T}(\omega)_B\hat{D}^\dagger(\beta_2)_{B}] \\
    &=\frac{1}{\pi^{2n}}\int d^{2n}\omega d^{2n}\beta_1d^{2n}\beta_2\lambda(\beta_2)g(\beta_1,\beta_2,r)e^{\zeta^*\cdot \omega^*-\zeta\cdot\omega}\delta(\omega+\beta_1)\delta(\omega^*+\beta_2) \\
    &=\frac{1}{\pi^{2n}}\int d^{2n}\omega\lambda(-\omega^*)g(-\omega,-\omega^*,r)e^{\zeta^*\cdot \omega^*-\zeta\cdot\omega}.
\end{align}
Hence, we finally obtain the probability of obtaining $\zeta$ from Bell measurement with an initial Bell state with finite squeezing:
\begin{align}
    p_{EA}(\zeta)&=\frac{1}{\pi^{2n}}\int d^{2n}\omega\lambda(-\omega^*)g(-\omega,-\omega^*,r)e^{\zeta^*\cdot \omega^*-\zeta\cdot\omega}
    =\frac{1}{\pi^{2n}}\int d^{2n}\omega\lambda(-\omega^*)e^{-e^{-2r}|\omega|^2}e^{\zeta^*\cdot \omega^*-\zeta\cdot\omega}.
\end{align}
By inverting the relation using Fourier transformation,
\begin{align}
    \int d^{2n}\zeta  p_{EA}(\zeta)e^{\zeta^\dagger\beta-\beta^\dagger\zeta}
    &=\frac{1}{\pi^{2n}}\int d^{2n}\omega\lambda(-\omega^*)g(-\omega,-\omega^*,r)e^{(\omega^*+\beta)\cdot\zeta^*-(\omega^*+\beta)^*\cdot\zeta} \\ 
    &=\lambda(\beta)g(\beta^*,\beta,r),
\end{align}
we obtain the relation between the characteristic function of the channel $\lambda(\beta)$ and the probability distribution of outcomes $p_{EA}(\zeta)$:
\begin{align}\label{eq:noiseless-g}
    \lambda(\beta)
    =\frac{1}{g(\beta^*,\beta,r)}\int d^{2n}\zeta  p_{EA}(\zeta)e^{\zeta^\dagger\beta-\beta^\dagger\zeta}
    =\exp(e^{-2r}|\beta|^2)\int d^{2n}\zeta  p_{EA}(\zeta)e^{\zeta^\dagger\beta-\beta^\dagger\zeta}.
\end{align}
The expression shows that by obtaining samples $\zeta$'s following the probability distribution using sampling in experiment and taking Fourier transformation, one can obtain the estimate of $\lambda(\beta)$.
Now, we show the number of samples $N\geq \frac{8}{\epsilon^2}\log\frac{4}{\delta}e^{2e^{-2r}|\beta|^2}=O(e^{2e^{-2r}|\beta|^2}\epsilon^{-2}\log\delta^{-1})$ suffices to have a good precision $\epsilon$ with a high probability $1-\delta$.
As observed above, for $N$ number of samples, $\{\zeta^{(i)}\}_{i=1}^N$, we set the estimator of $\lambda(\beta)$ to be $\tilde{\lambda}(\beta)=\frac{1}{N}\sum_{i=1}^N\exp(e^{-2r}|\beta|^2)e^{\zeta^{(i)\dagger}\beta-\beta^\dagger\zeta^{(i)}}$ for measurement outcome $\zeta$ and apply the Hoeffding bound for the estimator.
To do that, we find the bound for real part and imaginary part, respectively and combine them.
We first obtain two different probabilities' bound by the Hoeffding bound such that
\begin{align}
    \text{Pr}[|\tilde{\lambda}_r(\beta)-\lambda_r(\beta)|\leq \epsilon/2]\geq 1-2e^{-\frac{N\epsilon^2}{8}e^{-2e^{-2r}|\beta|^2}},~
    \text{Pr}[|\tilde{\lambda}_i(\beta)-\lambda_i(\beta)|\leq \epsilon/2]\geq 1-2e^{-\frac{N\epsilon^2}{8}e^{-2e^{-2r}|\beta|^2}},
\end{align}
where $\lambda_r=\text{Re}(\lambda)$ and $\lambda_i=\text{Im}(\lambda)$ and similar for $\tilde{\lambda}$.
Applying the union bound and the triangle inequality, we obtain
\begin{align}
    \text{Pr}[|\tilde{\lambda}(\beta)-\lambda(\beta)|\leq \epsilon]
    &\geq 1-4e^{-\frac{N\epsilon^2}{8}e^{-2e^{-2r}|\beta|^2}}.
\end{align}
Thus, if we choose the number of samples to be
\begin{align}
    N\geq \frac{8}{\epsilon^2}\log\frac{4}{\delta}e^{2e^{-2r}|\beta|^2}=O(e^{2e^{-2r}|\beta|^2}\epsilon^{-2}\log\delta^{-1})
\end{align}
the estimation error is upper-bounded by $\epsilon$ with high probability $1-\delta$.
This completes the proof of Theorem 1 in the main text.

Note that in an ideal case where the input squeezing parameter $r$ can be chosen to be arbitrarily large, the sample complexity can be reduced to $N=O(1/\epsilon^2)$ for any $\beta$.

\subsection{Entanglement-free (Vacuum+Heterodyne) schemes}\label{app:classical}
Now, let us consider the entanglement-free scheme with vacuum input and heterodyne detection (Vacuum+Heterodyne).
In general, denoting $\Pi_\phi$ as a POVM with an outcome $\phi$ and $|\phi_0\rangle\langle \phi_0|$ as an input state, the probability of a classical scheme is written as
\begin{align}\label{eq:prob_cl}
    p_{EF}(\phi)
    &=\Tr[\Pi_\phi\Lambda(|\phi_0\rangle\langle \phi_0|)] \\ 
    &=\frac{1}{\pi^{2n}}\int d^{2n}\alpha d^{2n}\beta \Tr[\hat{D}(\alpha)\Lambda(\hat{D}^\dagger(\beta))]\Tr[\Pi_\phi\hat{D}^\dagger(\alpha)]\Tr[|\phi_0\rangle\langle \phi_0|\hat{D}(\beta)] \\ 
    &=\frac{1}{\pi^{n}}\int d^{2n}\alpha d^{2n}\beta \lambda(\beta)\delta(\alpha-\beta)\Tr[\Pi_\phi\hat{D}^\dagger(\alpha)]\Tr[|\phi_0\rangle\langle \phi_0|\hat{D}(\beta)] \\ 
    &=\frac{1}{\pi^{n}}\int d^{2n}\beta \lambda(\beta)\Tr[\Pi_\phi\hat{D}^\dagger(\beta)]\Tr[|\phi_0\rangle\langle \phi_0|\hat{D}(\beta)].
\end{align}
For the Vacuum+Heterodyne scheme, we employ vacuum state input $|\phi_0\rangle=|0\rangle$ and heterodyne detection, whose POVM elements are described by projectors onto the (overcomplete) basis of coherent states, $|\zeta\rangle\langle\zeta|/\pi^n$, where $|\zeta\rangle$ is a coherent state with complex amplitude $\zeta\in\mathbb{C}^n$.
Note that such a scheme is informationally complete in the sense that it provides distinct probability distributions for different channels.
For this scheme, we can obtain the probability distribution
\begin{align}
    p_{VH}(\zeta)
    &=\frac{1}{\pi^{2n}}\int d^{2n}\alpha \lambda(\alpha)\Tr[|\zeta\rangle\langle \zeta|\hat{D}^\dagger(\alpha)]\Tr[|0\rangle\langle 0|\hat{D}(\alpha)] =\frac{1}{\pi^{2n}}\int d^{2n}\alpha \lambda(\alpha)e^{\alpha^\dagger\zeta-\zeta^\dagger\alpha}e^{-|\alpha|^2}.
\end{align}
Again, by inverting the probability distribution as
\begin{align}
    \int d^{2n}\zeta p_{VH}(\zeta)e^{\zeta^\dagger\beta-\beta^\dagger \zeta}
    =\frac{1}{\pi^{2n}}\int d^{2n}\zeta d^{2n}\alpha \lambda(\alpha)e^{\zeta^\dagger(\beta-\alpha)-(\beta-\alpha)^\dagger\zeta}e^{-|\alpha|^2}
    =\lambda(\beta)e^{-|\beta|^2},
\end{align}
we obtain the final relation between the measurement probability distribution and the characteristic function of the channel:
\begin{align}
    \lambda(\beta)=e^{|\beta|^2}\int d^{2n}\xi p_{VH}(\zeta)e^{\zeta^\dagger\beta-\beta^\dagger\zeta}.
\end{align}
It clearly shows the difference from the quantum strategies, which is the prefactor $e^{|\beta|^2}$.
Thus, similarly, after sampling $\zeta$ from experiments following $p_{EF}(\zeta)$ and averaging $e^{|\beta|^2}e^{\zeta^\dagger\beta-\beta^\dagger\zeta}$ over the samples, we obtain the estimate of $\lambda(\beta)$.
As for the entanglement-assisted case, for $N$ samples, $\{\zeta^{(i)}\}_{i=1}^N$, by setting the estimator $\tilde{\lambda}(\beta)=\frac{1}{N}\sum_{i=1}^N e^{|\beta|^2}e^{\zeta^{(i)\dagger}\beta-\beta^\dagger\zeta^{(i)}}$ and using the Hoeffding bound, we obtain
\begin{align}
    \text{Pr}[|\tilde{\lambda}(\beta)-\lambda(\beta)|\leq \epsilon]
    &\geq 1-4e^{-\frac{N\epsilon^2}{8}e^{-2|\beta|^2}}.
\end{align}
Thus, in this case, it indicates that the sufficient number of samples is
\begin{align}
    N\geq \frac{8}{\epsilon^2}\log\frac{4}{\delta}e^{2|\beta|^2}=O(e^{2|\beta|^2}\epsilon^{-2}\log\delta^{-1})
\end{align}
for the estimation error to be upper-bounded by $\epsilon$ with high probability $1-\delta$.
It clearly shows the significant difference of the sample complexity from the entanglement-assisted case.

\subsection{Entanglement-assisted scheme with imperfection}\label{app:combine}
We now consider the effect of imperfections and prove Theorem 3 in the main text.
More specifically, we consider the cases where photon loss occurs before and after applying the random displacement channel we want to learn and a regularized Bell measurement.
Here, the photon loss before and after the random displacement channel models an imperfect input state preparation and imperfect Bell measurement.
On the other hand, we introduce parameter $s$ to regularize the Bell measurement POVM by a general-dyne measurement POVM, where we recover the perfect Bell measurement by taking $s \to \infty$.
By considering the regularized Bell measurement POVM, we assume the same condition as the lower bound in Sec.~\ref{app:efree} in that the measurement POVM is normalizable, i.e., its norm is finite.

Let us first consider the effect of the loss channel on a single-mode displacement operator.
Using the equivalent description of a loss channel $\mathcal{L}$ by a beam splitter interaction with a vacuum environment, we can show that
\begin{align}
    \mathcal{L}[\hat{D}^\dagger(\alpha)_S]
    &=\Tr_E[U_T(\hat{D}^\dagger(\alpha)_S\otimes |0\rangle\langle 0|_E)U_T^\dagger] \\ 
    &=\int \frac{d^{2}z}{\pi^n}e^{-\frac{1}{2}|z|^2}\Tr_E[U_T\hat{D}^\dagger(\alpha)_S\otimes \hat{D}^\dagger(z)_EU_T^\dagger] \\ 
    &=\int \frac{d^{2}z}{\pi^n}e^{-\frac{1}{2}|z|^2}\Tr_E[\hat{D}^\dagger(\sqrt{T}\alpha-\sqrt{1-T}z)_S\otimes \hat{D}^\dagger(\sqrt{T}z+\sqrt{1-T}\alpha)_E] \\ 
    &=T^{-1}e^{-\frac{1-T}{2T}|\alpha|^2}\hat{D}^\dagger(\alpha/\sqrt{T})_S,
\end{align}
where $\hat{U}_T$ is the beam splitter interaction with the environment with the transmission rate $T$, and thus $1-T$ is the loss rate.
Here, we used the identity~\cite{ferraro2005gaussian}
\begin{align}
    |0\rangle\langle 0|=\int \frac{d^{2n}z}{\pi^n}e^{-\frac{1}{2}|z|^2}\hat{D}^\dagger(z).
\end{align}

Recall that the input state is two-mode squeezed states with a finite squeezing parameter, which is written as
\begin{align}
    |\tilde{\Psi}(r)\rangle\langle \tilde{\Psi}(r)|
    =\frac{1}{\pi^{2n}}\int d^{2n}\omega_1 d^{2n}\omega_2g(w_1,w_2,r)\hat{D}^\dagger(\omega_1)\otimes \hat{D}^\dagger(\omega_2).
\end{align}
Let us first study how the displacement operator transforms over the channels.

First, after a loss channel with loss rate $1-T_b$, the random displacement channel $\Lambda$ and another loss channel with loss rate $1-T_a$, an $n$-mode displacement operator transforms as
\begin{align}
    &\mathcal{L}_{AB}^{T_a}(\mathcal{I}_A\otimes \Lambda_B)\mathcal{L}_{AB}^{T_b}(\hat{D}^\dagger(\omega_1)_A\otimes \hat{D}^\dagger(\omega_2)_B) \\ 
    &=T_b^{-2n}e^{-\frac{1-T_b}{2T_b}(|\omega_1|^2+|\omega_2|^2)}\mathcal{L}_{AB}^{T_a}(\mathcal{I}_A\otimes \Lambda_B)(\hat{D}^\dagger(\omega_1/\sqrt{T_b})_A\otimes \hat{D}^\dagger(\omega_2/\sqrt{T_b})_B) \\ 
    &=T_b^{-2n}e^{-\frac{1-T_b}{2T_b}(|\omega_1|^2+|\omega_2|^2)}\lambda(\omega_2/\sqrt{T_b})\mathcal{L}_{AB}^{T_a}(\hat{D}^\dagger(\omega_1/\sqrt{T_b})_A\otimes \hat{D}^\dagger(\omega_2/\sqrt{T_b})_B) \\ 
    &=(T_bT_a)^{-2n}e^{-\frac{1-T_b}{2T_b}(|\omega_1|^2+|\omega_2|^2)}e^{-\frac{1-T_a}{2T_aT_b}(|\omega_1|^2+|\omega_2|^2)}\lambda(\omega_2/\sqrt{T_b})(\hat{D}^\dagger(\omega_1/\sqrt{T_bT_a})_A\otimes \hat{D}^\dagger(\omega_2/\sqrt{T_bT_a})_B).
\end{align}
Now we implement the regularized Bell measurement.
Recall that the perfect Bell measurement can be conducted by applying a 50:50 beam splitter and then performing homodyne detection.
Here, we will regularize the homodyne detection by general-dyne detection and tracing out some quadratures.

After 50:50 beam splitters $\hat{U}_\text{BS}$, the displacement operators transform as
\begin{align}
    \hat{D}^\dagger(\omega_1/\sqrt{T_bT_a})_A\otimes \hat{D}^\dagger(\omega_2/\sqrt{T_bT_a})_B
    \to \hat{D}^\dagger\left(\frac{\omega_1-\omega_2}{\sqrt{2T_bT_a}}\right)_A\otimes \hat{D}^\dagger\left(\frac{\omega_1+\omega_2}{\sqrt{2T_bT_a}}\right)_B.
\end{align}
We then perform measurements described by the following POVM:
\begin{align}
    \bigl\{\hat\Pi^A_{\alpha}(-s)\otimes\hat\Pi^B_{\beta}(s)\bigr\}_{\alpha,\beta}~~~\text{where}~~~\hat\Pi_{\gamma}(s) \coleq \frac{1}{\pi^n}\hat D(\gamma)\ketbra{s}{s}\hat D^\dagger(\gamma),
\end{align}
where $s\ge0$ is the squeezing parameter for the Bell measurement.
We note that this measurement corresponds to a special type of general-dyne measurement \cite{serafini2017quantum} and that when $s\to \infty$, we recover the Bell measurement studied in Sec.~\ref{app:bell_finite}.
Then, the output probability is written as
\begin{align}
    q(\alpha,\beta)
    \coleq \frac{1}{\pi^{2n}}\int d^{2n}\omega g(\omega_1,\omega_2,r)\Tr[\hat\Pi^A_{\alpha}(-s)\otimes\hat\Pi^B_{\beta}(s)\hat{U}_{BS}\left(\mathcal{L}_{AB}^{T_a}(\mathcal{I}_A\otimes \Lambda_B)\mathcal{L}_{AB}^{T_b}(\hat{D}^\dagger(\omega_1)_A\otimes \hat{D}^\dagger(\omega_2)_B)\right)\hat{U}_{BS}^\dagger].
\end{align}
Here, we have
\begin{align}
    \Tr[\hat{\Pi}_\alpha(-s)\hat{D}^\dagger(\omega)]
    &=\frac{1}{\pi^n}\langle -s|\hat{D}^\dagger(\alpha)\hat{D}^\dagger(\omega)\hat{D}(\alpha)|-s\rangle=\frac{1}{\pi^n}e^{-\frac{1}{2}(|\omega|^2\cosh{2s}+(\omega^2+\omega^{*2})\sinh{2s}/2)} e^{\omega^\dagger\alpha-\alpha^\dagger\omega}, \\
    \Tr[\hat{\Pi}_\beta(s)\hat{D}^\dagger(\omega)]
    &=\frac{1}{\pi^n}e^{-\frac{1}{2}(|\omega|^2\cosh{2s}-(\omega^2+\omega^{*2})\sinh{2s}/2)} e^{\omega^\dagger\beta-\beta^\dagger\omega},
\end{align}
and
\begin{align}
    &\Tr[\hat{\Pi}_\alpha(-s)\hat{D}^\dagger\left(\frac{\omega_1-\omega_2}{\sqrt{2T_aT_b}}\right)]\Tr[\hat{\Pi}_\beta(s)\hat{D}^\dagger\left(\frac{\omega_1+\omega_2}{\sqrt{2T_aT_b}}\right)] \\
    &=\frac{1}{\pi^{2n}}e^{-\frac{1}{2}\left[(|\omega_1|^2+|\omega_2|^2)\frac{\cosh{2s}}{T_aT_b}-(\omega_1\cdot \omega_2+\omega_1^*\cdot \omega_2^*)\frac{\sinh{2s}}{T_aT_b}\right]}\exp\left[\omega_1^*\left(\frac{\alpha+\beta}{\sqrt{2T_aT_b}}\right)+\omega_2^*\left(\frac{\beta-\alpha}{\sqrt{2T_aT_b}}\right)-c.c.\right] \\
    &=\frac{1}{\pi^{2n}}g(\omega_1/\sqrt{T_aT_b},\omega_2/\sqrt{T_aT_b},s)\exp\left[\omega_1^\dagger\left(\frac{\alpha+\beta}{\sqrt{2T_aT_b}}\right)+\omega_2^\dagger\left(\frac{\beta-\alpha}{\sqrt{2T_aT_b}}\right)-c.c.\right].
\end{align}

We now trace out one of two quadratures for each mode.
To do that, we take integral over the imaginary part of $\alpha$ and the real part of $\beta$.
Here, we define $\alpha\coleq \alpha_r+i\alpha_i$ and $\beta\coleq \beta_r+i\beta_i$.
If we take the integral, the relevant part reduces to 
\begin{align}
    \frac{1}{\pi^n}\int d^n\alpha_i \exp\left[\alpha \left(\frac{\omega_1^*-\omega_2^*}{\sqrt{2T_aT_b}}\right)-\alpha^* \left(\frac{\omega_1-\omega_2}{\sqrt{2T_aT_b}}\right)\right]
    &=\delta(\Re (\omega_1-\omega_2)/\sqrt{2T_aT_b}) e^{-i\sqrt{2}\Im (\omega_1-\omega_2)\alpha_r/\sqrt{T_aT_b}}, \\
    \frac{1}{\pi^n}\int d^n\beta_r \exp\left[\beta \left(\frac{\omega_1^*+\omega_2^*}{\sqrt{2T_aT_b}}\right)-\beta^* \left(\frac{\omega_1+\omega_2}{\sqrt{2T_aT_b}}\right)\right]
    &=\delta(\Im(\omega_1+\omega_2)/\sqrt{2T_aT_b})e^{i\sqrt{2}\Re(\omega_1+\omega_2)\beta_i/\sqrt{T_aT_b}},
\end{align}
where the delta function gives us $\omega_2=\omega_1^*$.
Thus,
\begin{align}
    &\int d^n\alpha_id^n\beta_r\Tr[\hat{\Pi}_\alpha(-s)\hat{D}^\dagger\left(\frac{\omega_1-\omega_2}{\sqrt{2T_aT_b}}\right)]\Tr[\hat{\Pi}_\beta(s)\hat{D}^\dagger\left(\frac{\omega_1+\omega_2}{\sqrt{2T_aT_b}}\right)] \\ 
    &=g(\omega_1/\sqrt{T_aT_b},\omega_2/\sqrt{T_aT_b},s)\delta\left(\frac{\omega_1-\omega_2^*}{\sqrt{2T_aT_b}}\right)e^{-i\sqrt{2}\Im (\omega_1-\omega_2) \alpha_r/\sqrt{T_aT_b}}e^{i\sqrt{2}\Re(\omega_1+\omega_2)\beta_i/\sqrt{T_aT_b}}.
\end{align}
Thus, the output probability is, by defining the measurement output variable as $\xi=-\alpha_r+i\beta_i$, given by
\begin{align}
    &q(\xi) \\ 
    &=\int d^n\alpha_id^n\beta_rq(\alpha,\beta) \\
    &=\frac{1}{\pi^{2n}}\int d^n\alpha_id^n\beta_rd^{2n}\omega_1d^{2n}\omega_2 (T_bT_a)^{-n}e^{-(\frac{1-T_b}{2T_b}+\frac{1-T_a}{2T_aT_b})(|\omega_1|^2+|\omega_2|^2)}g(\omega_1,\omega_2,r)\lambda(\omega_2/\sqrt{T_b}) \nonumber \\ 
    &~~~~~~~~~~~~~~~~\times \Tr[\hat{\Pi}_\alpha(-s)\hat{D}^\dagger\left(\frac{\omega_1-\omega_2}{\sqrt{2T_aT_b}}\right)]\Tr[\hat{\Pi}_\beta(s)\hat{D}^\dagger\left(\frac{\omega_1+\omega_2}{\sqrt{2T_aT_b}}\right)] \\ 
    &=\frac{1}{\pi^{2n}}\int d^{2n}\omega_1d^{2n}\omega_2 (T_bT_a)^{-2n}e^{-(\frac{1-T_b}{2T_b}+\frac{1-T_a}{2T_aT_b})(|\omega_1|^2+|\omega_2|^2)}\lambda(\omega_2/\sqrt{T_b})g(\omega_1,\omega_2^*,r)g(\omega_1/\sqrt{T_aT_b},\omega_2/\sqrt{T_aT_b},s) \nonumber \\ 
    &~~~~~~~~~~~~~~~~\times \delta\left(\frac{\omega_1-\omega_2^*}{\sqrt{2T_aT_b}}\right) e^{-i\sqrt{2}\Im (\omega_1-\omega_2)\alpha_r/\sqrt{T_aT_b}}e^{i\sqrt{2}\Re(\omega_1+\omega_2)\beta_i/\sqrt{T_aT_b}} \\
    &=\left(\frac{2}{T_aT_b}\right)^n\frac{1}{\pi^{2n}}\int d^{2n}\omega e^{-(\frac{1-T_b}{T_b}+\frac{1-T_a}{T_aT_b})|\omega|^2}\lambda(\omega^*/\sqrt{T_b})g(\omega,\omega^*, r)g(\omega/\sqrt{T_aT_b},\omega^*/\sqrt{T_aT_b},s)e^{\sqrt{2}(\xi\cdot\omega-\omega^*\cdot\xi^*)/\sqrt{T_aT_b}} \\
    &=\left(\frac{2}{T_aT_b}\right)^n\frac{1}{\pi^{2n}}\int d^{2n}\omega e^{-(\frac{1-T_b}{T_b}+\frac{1-T_a}{T_aT_b})|\omega|^2}\lambda(\omega^*/\sqrt{T_b})e^{-(e^{-2r}+e^{-2s}/T_aT_b)|\omega|^2}e^{\sqrt{2}(\xi\cdot\omega-\omega^*\cdot\xi^*)/\sqrt{T_aT_b}} \\
    &=\left(\frac{2}{T_a}\right)^n\frac{1}{\pi^{2n}}\int d^{2n}\omega e^{-((1-T_b)+\frac{1-T_a}{T_a})|\omega|^2}\lambda(\omega^*)e^{\sqrt{2}(\xi\cdot\omega-\omega^*\cdot\xi^*)/\sqrt{T_a}}.
\end{align}
Here, for consistency with Sec.~\ref{app:bell_finite}, we rescale $\sqrt{2}\xi=\zeta$ and define $p_{loss}(\zeta)$ such that
\begin{align}
    p_{loss}(\zeta)=\left(\frac{1}{T_a}\right)^n\frac{1}{\pi^{2n}}\int d^{2n}\omega e^{-((1-T_b)+\frac{1-T_a}{T_a})|\omega|^2}\lambda(\omega^*)e^{-(T_be^{-2r}+e^{-2s}/T_a)|\omega|^2}e^{(\omega\cdot\zeta-\omega^*\cdot\zeta^*)/\sqrt{T_a}},
\end{align}
where $2^n$ factor is canceled because of the relation $2^n d^{2n}\xi=d^{2n}\zeta$ (This rescaling is because the convention of Bell measurement outcome $\zeta$ in Sec.~\ref{app:bell_finite} is different from $\xi$ in this section by $\sqrt{2}$ factor.).
Therefore, after Fourier transformation, we obtain
\begin{align}
    &\int d^{2n}\zeta p_{loss}(\zeta) e^{(\zeta^\dagger\beta-\beta^\dagger\zeta)/\sqrt{T_a}}
    =\lambda(\beta)e^{-(T_be^{-2r}+e^{-2s}/T_a)|\beta|^2}e^{-(1-T_b)|\beta|^2}e^{-\frac{1-T_a}{T_a}|\beta|^2}.
\end{align}
Consequently, the characteristic function is written by the probability distribution as
\begin{align}
    \lambda(\beta)&=e^{(T_be^{-2r}+e^{-2s}/T_a)|\beta|^2}e^{(1-T_b)|\beta|^2}e^{\frac{1-T_a}{T_a}|\beta|^2}\int d^{2n}\zeta p_{loss}(\zeta) e^{(\zeta^\dagger\beta-\beta^\dagger\zeta)/\sqrt{T_a}},\\
    &= e^{e^{-2r_\mr{eff}}|\beta|^2}\int d^{2n}\zeta p_{loss}(\zeta) e^{(\zeta^\dagger\beta-\beta^\dagger\zeta)/\sqrt{T_a}}.
\end{align}
where we defined an effective squeezing parameter $r_\mr{eff}$ due to all kinds of imperfections via 
\begin{align}
    e^{-2r_\text{eff}}\coleq (T_be^{-2r}+T_a^{-1}e^{-2s})+(1-T_b)+\frac{1-T_a}{T_a}.
\end{align}
In order to estimate any $\lambda(\beta)$, one simply obtains $N$ samples $\{\zeta^{(i)}\}_{i=1}^N$ from $p(\zeta)$ and set the estimator to be $\tilde{\lambda}(\beta)\coleq \frac{1}{N}\sum_{i=1}^Ne^{e^{-2r_\mr{eff}}|\beta|^2}e^{(\zeta^{(i)\dagger}\beta-\beta^\dagger\zeta^{(i)})/\sqrt{T_a}}$. According to the Hoeffding's inequality as the ideal case, averaging over $N\geq 8/\epsilon^2 \log(4/\delta) e^{2e^{-2r_\mr{eff}}|\beta|^2}$ copies is sufficient to estimate $\lambda(\beta)$ to $\epsilon$ additive error with high probability.

Meanwhile, the effects of imperfections are thus the envelope of Fourier transforms.
Especially when $s\to \infty$, the effective squeezing parameter under loss is given by
\begin{align}
    r_\text{eff}=-\frac{1}{2}\log\left(T_be^{-2r}+(1-T_b)+\frac{1-T_a}{T_a}\right).
\end{align}
This completes the proof of Theorem 3 in the main text.

To be clearer, for photon loss before the channel without any other imperfections, i.e., $s\to \infty$ and $T_a=1$, the envelope is given by
\begin{align}
    e^{[T_b e^{-2r}+(1-T_b)]|\beta|^2},
\end{align}
and for photon loss after the channel without other imperfections, the envelope is given by
\begin{align}
    e^{[e^{-2r}+(1-T_a)/T_a]|\beta|^2}.
\end{align}

\subsection{Discussion on more general input states}
In this section, we study more general sources of noise other than finite squeezing and photon loss. To begin with, consider the case where we use an arbitrary input state while the CV Bell measurement is still employed.
To this end, note that Eq.~\eqref{eq:noiseless-g} from Sec.~\ref{app:bell_finite} does not use any special properties of TMSV, and actually holds for any $2n$-mode input state, \textit{i.e.},
\begin{align}
    \lambda(\beta)
    =\frac{1}{g_{\hat{\rho}}(\beta^*,\beta)}\int d^{2n}\zeta  p_{EA}(\zeta)e^{\zeta^\dagger\beta-\beta^\dagger\zeta},
\end{align}
where $g$ is the characteristic function of the input state $\hat{\rho}$:
\begin{align}
    g_{\hat{\rho}}(w_1,w_2)=\Tr[\hat{\rho}\hat{D}(\omega_1)\otimes \hat{D}(\omega_2)].
\end{align}
The fact that the same relation holds by replacing the $g$ function properly indicates that for different types of input states, we still have a very similar form of an unbiased estimator for $N$ samples:
\begin{align}
    \tilde{\lambda}(\beta)=\frac{1}{N}\sum_{i=1}^N \frac{1}{g_{\hat{\rho}}(\beta^*,\beta)}e^{\zeta^{(i)\dagger}\beta-\beta^\dagger\zeta^{(i)}}.
\end{align}
It implies that the sampling complexity is determined by the function $g_{\hat{\rho}}(\beta^*,\beta)$.
More specifically, by the Hoeffding inequality, the number of samples to achieve an error $\epsilon$ with high probability $1-\delta$ is given by
\begin{align}
    N=O(|g_{\hat{\rho}}(\beta^*,\beta)|^{-2}\epsilon^{-2}\log\delta^{-1}).
\end{align}
For example, for TMSV states, this function reduces to
\begin{align}
    g_{\tilde{\Psi}}(\beta^*,\beta)=\exp(-e^{-2r}|\beta|^2).
\end{align}
Therefore, as long as the function $g$ of the input state is sufficiently large for the $\beta$ of interest, we can still expect the scheme to be sample efficient.

\begin{figure*}[t]
	\centering
        \includegraphics[width=0.97\linewidth]{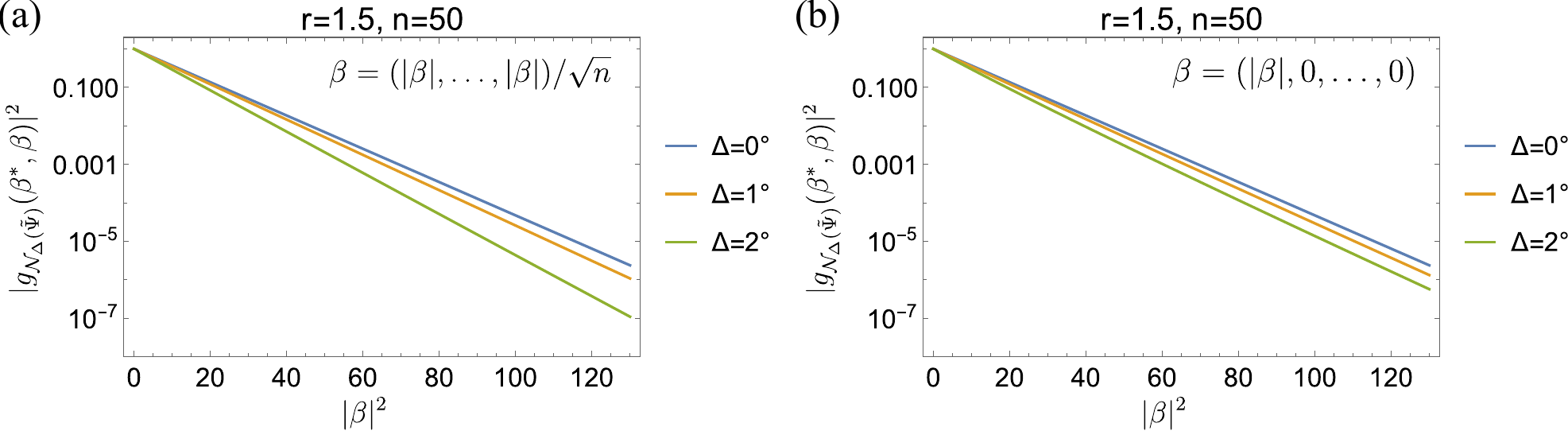}
        \caption{Effect of phase diffusion on the squared characteristic function $|g_{\mathcal{N}_\Delta(\hat{\rho})}(\beta^*,\beta)|^2$, where we choose a specific form of $\beta\in\mathbb{C}^n$ as (a) $\beta\coleq(|\beta|/\sqrt{n},\dots,|\beta|/\sqrt{n})$ and (b) $\beta\coleq(|\beta|,0,\dots,0)$ for two extreme cases with a given $|\beta|^2$. We fix the squeezing parameter $r=1.5$ of the input TMSV states and set the number of modes $n=50$. For the standard deviation $\Delta=1^{\circ}$ of phase noise, following Gaussian distributions, one may observe that the characteristic function is almost identical to the noiseless case~($\Delta=0^{\circ}$). For the standard deviation $\Delta=2^{\circ}$ as well, the effect is not very significant.
        }
	\label{fig:phase}
\end{figure*}

Such a general form enables us to analyze the effect of general noise on input states.
Let us again focus on TMSV states but assume a noise channel $\mathcal{N}$.
Then, the characteristic function $g$ of the noisy TMSV states can be written as
\begin{align}
    g_{\mathcal{N}(\tilde{\Psi})}(w_1,w_2)=\Tr[\mathcal{N}(|\tilde{\Psi}(r)\rangle\langle \tilde{\Psi}(r)|)\hat{D}(\omega_1)\otimes \hat{D}(\omega_2)].
\end{align}
As discussed, it suffices to analyze how the characteristic function changes by noise to ensure that a significant advantage is still maintained for noisy states.
In typical experiments, while the CV Bell measurement noise can be modeled by photon loss, as we considered already in the previous section, other types of noise may exist in the TMSV state preparation procedure.
An example is phase diffusion, which can be modeled by a photon-number-dependent random phase following a Gaussian distribution.
For $2n$-mode state input, we can write the noise channel as
\begin{align}
    &\mathcal{N}_\Delta(\hat{\rho}) \\
    &=\int d^{n}\phi_Ad^{n}\phi_B\frac{e^{-\frac{|\phi_A|^2+|\phi_B|^2}{2\Delta^2}}}{(2\pi\Delta^2)^n} e^{-i\phi_A\cdot \hat{n}_A-i\phi_B\cdot \hat{n}_B}\hat{\rho} e^{i\phi_A\cdot \hat{n}_A+i\phi_B\cdot \hat{n}_B} \\
    &=\frac{1}{\pi^{2n}}\int d^{2n}\omega_1d^{2n}\omega_2 g_{\hat{\rho}}(\omega_1,\omega_2)\int d^{n}\phi_Ad^{n}\phi_B\frac{e^{-\frac{|\phi_A|^2+|\phi_B|^2}{2\Delta^2}}}{(2\pi\Delta^2)^n} e^{-i\phi_A\cdot \hat{n}_A-i\phi_B\cdot \hat{n}_B}[\hat{D}^\dagger(\omega_1)\otimes \hat{D}^\dagger(\omega_2)] e^{i\phi_A\cdot \hat{n}_A+i\phi_B\cdot \hat{n}_B}
    ,
\end{align}
where $\phi_A$ and $\phi_B$ are $n$-dimensional real vectors and $\hat{n}_A$ and $\hat{n}_B$ are photon number operator vectors for $A$ and $B$ parts, respectively.
Then, noting that
\begin{align}
    e^{-i\phi_A\cdot \hat{n}_A-i\phi_B\cdot \hat{n}_B}[\hat{D}^\dagger(\omega_1)\otimes \hat{D}^\dagger(\omega_2)] e^{i\phi_A\cdot \hat{n}_A+i\phi_B\cdot \hat{n}_B}
    =\hat{D}^\dagger(\omega_1 e^{-i\phi_A})\otimes \hat{D}^\dagger(\omega_2 e^{-i\phi_B}),
\end{align}
where $\omega_1 e^{-i\phi_A}$ and $\omega_2 e^{-i\phi_B}$ are interpreted as vectors obtained by an elementwise product, the corresponding $g$ function for TMSV states is written as
\begin{align}
    &g_{\mathcal{N}(\tilde{\Psi})}(w_1,w_2) \\
    &=\Tr[\mathcal{N}_\Delta(|\tilde{\Psi}(r)\rangle\langle \tilde{\Psi}(r)|)\hat{D}(\omega_1)\otimes \hat{D}(\omega_2)] \\
    &=
    \frac{1}{\pi^{2n}}\int d^{2n}\beta_1d^{2n}\beta_2 g_{\tilde{\Psi}}(\beta_1,\beta_2)\int d^{n}\phi_Ad^{n}\phi_B\frac{e^{-\frac{|\phi_A|^2+|\phi_B|^2}{2\Delta^2}}}{(2\pi\Delta^2)^n} 
    \Tr[\hat{D}^\dagger(\beta_1 e^{-i\phi_A})\otimes \hat{D}^\dagger(\beta_2 e^{-i\phi_B})\hat{D}(\omega_1)\otimes \hat{D}(\omega_2)] \\
    &=
    \int d^{2n}\beta_1d^{2n}\beta_2 g_{\tilde{\Psi}}(\beta_1,\beta_2)\int d^{n}\phi_Ad^{n}\phi_B\frac{e^{-\frac{|\phi_A|^2+|\phi_B|^2}{2\Delta^2}}}{(2\pi\Delta^2)^n} 
    \delta(\omega_1-\beta_1e^{-i\phi_A})\delta(\omega_2-\beta_2e^{-i\phi_B}) \\
    &=
    \int d^{n}\phi_Ad^{n}\phi_B\frac{e^{-\frac{|\phi_A|^2+|\phi_B|^2}{2\Delta^2}}}{(2\pi\Delta^2)^n} 
    g_{\tilde{\Psi}}(\omega_1 e^{i\phi_A},\omega_2 e^{i\phi_B}).
\end{align}
Thus, the effect of phase noise is to transform the $g$ function as a mixture with random phases.
We present examples to illustrate the effect of the phase noise on the sample complexity in Fig.~\ref{fig:phase}.
We have chosen the parameters $\beta\in\mathbb{C}^n$ of two extreme cases as $\beta\coleq (|\beta|/\sqrt{n},\dots,|\beta|/\sqrt{n})$ and $\beta\coleq (|\beta|,0,\dots,0)$.
Recall that the typical choice of the regime of interest in the main text is $|\beta|^2\leq \kappa n$; here, the range $|\beta|^2\in[0,130]$ in the figure covers up to $\kappa=2.5$ for $n=50$.
We see that the advantages of the entanglement-assisted scheme look robust against small-phase diffusion noise.

\subsection{Entanglement-assisted scheme with crosstalk}
In this subsection, we analyze the effect of crosstalk on measurement.
More specifically, we consider the crosstalk for the CV Bell measurement that occurs due to the imprecision of the interference phase between the two modes before the measurement.
This effect can be described by applying an additional beam splitter with nonzero angle $\theta$, $\hat{U}_\text{ct}$, before the 50:50 beam splitter for each Bell measurement.
Hence, the output probability of the TMSV+BM scheme under the effect of crosstalk is modified as
\begin{align}
    &p_\text{ct}(\zeta) \\
    &=\frac{1}{\pi^{n}}\Tr[(|\Psi(\zeta)\rangle\langle \Psi(\zeta)|_{AB})\hat{U}_\text{ct}(\mathcal{I}_A\otimes \Lambda_{B})(|\tilde{\Psi}\rangle\langle \tilde{\Psi}|_{AB})\hat{U}_\text{ct}^\dagger] \\ 
    &=\frac{1}{\pi^{4n}}\Tr[\int d^{2n}\omega d^{2n}\omega_1 d^{2n}\omega_2 g(w_1,w_2,r) e^{\zeta^*\cdot \omega^*-\zeta\cdot\omega}[\hat{D}^\dagger(\omega)\otimes \hat{D}^\text{T}(\omega)]\hat{U}_\text{ct}(\mathcal{I}_A\otimes \Lambda_{B})(\hat{D}^\dagger(\omega_1)\otimes \hat{D}^\dagger(\omega_2))\hat{U}_\text{ct}^\dagger] \\
    &=\frac{1}{\pi^{4n}}\int d^{2n}\omega d^{2n}\omega_1 d^{2n}\omega_2 \lambda(\omega_2)g(w_1,w_2,r) e^{\zeta^*\cdot \omega^*-\zeta\cdot\omega}\Tr[[\hat{D}^\dagger(\omega)\otimes \hat{D}^\text{T}(\omega)]\hat{U}_\text{ct}[\hat{D}^\dagger(\omega_1)\otimes \hat{D}^\dagger(\omega_2)]\hat{U}_\text{ct}^\dagger] \\ 
    &=\frac{1}{\pi^{2n}}\int d^{2n}\omega d^{2n}\omega_1 d^{2n}\omega_2 \lambda(\omega_2)g(w_1,w_2,r) e^{\zeta^*\cdot \omega^*-\zeta\cdot\omega}\delta(\omega+\omega_1\cos\theta+\omega_2\sin\theta)\delta(\omega^*+\omega_2\cos\theta-\omega_1\sin\theta) \\ 
    &=\frac{1}{\pi^{2n}}\int d^{2n}\omega_1 d^{2n}\omega_2 \lambda(\omega_2)g(w_1,w_2,r) e^{\zeta\cdot(\omega_1\cos\theta+\omega_2\sin\theta)-\zeta^*\cdot (\omega_1\cos\theta+\omega_2\sin\theta)^*} \\ 
    &\quad \quad \times \delta(-\omega_1^*\cos\theta-\omega_2^*\sin\theta+\omega_2\cos\theta-\omega_1\sin\theta) \\ 
    &=\frac{1}{\pi^{2n}}\int d^{2n}\omega \lambda(\omega)g(w^\theta,w,r) e^{\zeta\cdot(\omega^\theta\cos\theta+\omega\sin\theta)-\zeta^*\cdot (\omega^\theta\cos\theta+\omega\sin\theta)^*} \\ 
    &=\frac{1}{\pi^{2n}}\int d^{2n}\omega \lambda(\omega)g(w^\theta,w,r) e^{\frac{2i \zeta_i\cdot \omega_r}{\sin\theta+\cos\theta}+\frac{2i\zeta_r\cdot\omega_i}{\sin\theta-\cos\theta}}.
\end{align}
where we defined
\begin{align}
    \Re\omega^\theta=\frac{\cos\theta-\sin\theta}{\sin\theta+\cos\theta}\Re\omega,~~~
    \Im\omega^\theta=\frac{\sin\theta+\cos\theta}{\sin\theta-\cos\theta}\Im\omega,
\end{align}
which satisfies
\begin{align}
    \omega^\theta\cos\theta+\omega\sin\theta=\frac{\Re\omega}{\sin\theta+\cos\theta}+i\frac{\Im\omega}{\sin\theta-\cos\theta}.
\end{align}

\begin{figure*}[t]
	\centering
        \includegraphics[width=0.97\linewidth]{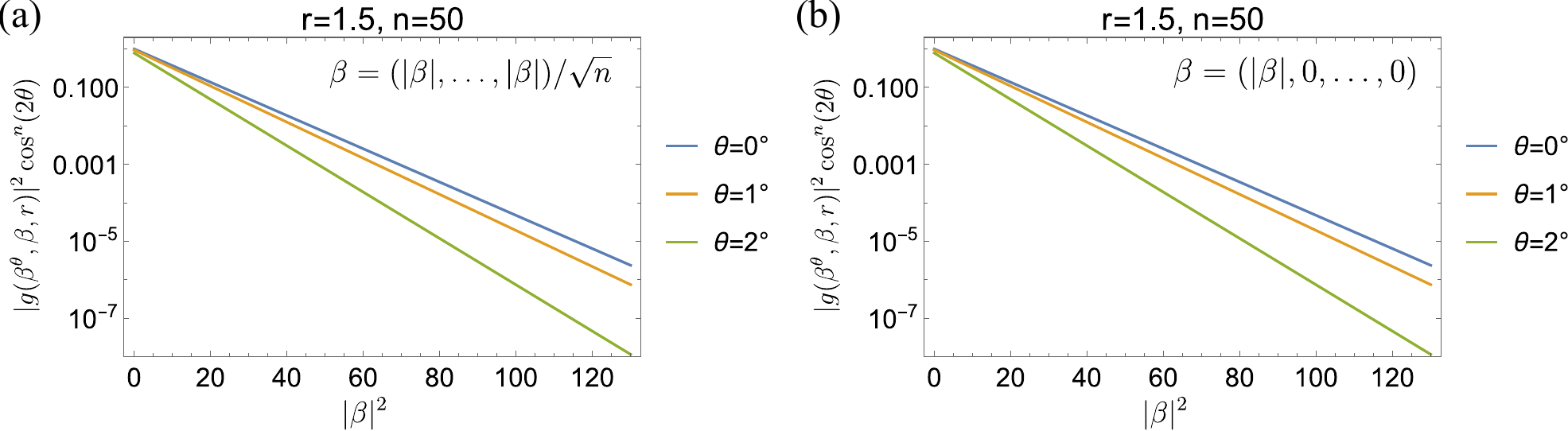}
        \caption{Effect of crosstalk on the squared characteristic function $|g(\beta^\theta,\beta,r)|^2\cos^{2n}(2\theta)$, where we choose a specific form of $\beta\in\mathbb{C}^n$ as (a) $\beta\coleq(|\beta|/\sqrt{n},\dots,|\beta|/\sqrt{n})$ and (b) $\beta\coleq(|\beta|,0,\dots,0)$ for two extreme cases with a given $|\beta|^2$. We fix the squeezing parameter $r=1.5$ of the input TMSV states and set the number of modes $n=50$. $\theta$ characterizes the crosstalk rate. One may observe that the sample overhead, the inverse of the $y$-axis, does not significantly change when the crosstalk of Bell measurement is not large.
        }
	\label{fig:crosstalk}
\end{figure*}

By taking the Fourier transform to find an unbiased estimator of $\lambda(\beta)$ as before, we obtain
\begin{align}
    &\int d^{2n}\zeta p_\text{ct}(\zeta)e^{-\frac{2i\zeta_i\beta_r}{\sin\theta+\cos\theta}-\frac{2i\zeta_r\beta_i}{\sin\theta-\cos\theta}} \\
    &=\frac{1}{\pi^{2n}}\int d^{2n}\omega \lambda(\omega)g(w^\theta,w,r) e^{\frac{2i \zeta_i\cdot (\omega_r-\beta_r)}{\sin\theta+\cos\theta}+\frac{2i\zeta_r\cdot(\omega_i-\beta_i)}{\sin\theta-\cos\theta}} \\ 
    &=\int d^{2n}\omega \lambda(\omega)g(w^\theta,w,r) \delta(\frac{\omega_r-\beta_r}{\sin\theta+\cos\theta})\delta(\frac{\omega_i-\beta_i}{\sin\theta-\cos\theta}) \\ 
    &=\lambda(\beta)g(\beta^\theta,\beta,r)\cos^{n}(2\theta),
\end{align}
which gives the relation between the characteristic function $\lambda(\beta)$ and the output probability distributuon
\begin{align}
    \lambda(\beta)=\frac{1}{g(\beta^\theta,\beta,r)\cos^{n}(2\theta)}\int d^{2n}\zeta p_\text{ct}(\zeta)e^{-\frac{2i\zeta_i\beta_r}{\sin\theta+\cos\theta}-\frac{2i\zeta_r\beta_i}{\sin\theta-\cos\theta}}.
\end{align}
One may notice that the estimator has to be appropriately adjusted following the ratio of $\theta$ that characterizes the crosstalk rate.

With the same logic as previous cases using Hoeffding inequality, the above form implies that the number of copies $N$ to achieve an error $\epsilon$ with high probability $1-\delta$ is given by
\begin{align}
    N=O(|g(\beta^\theta,\beta,r)|^{-2}\cos^{-2n}(2\theta) \epsilon^{-2}\log\delta^{-1}).
\end{align}

To analyze the effect of the crosstalk more quantitatively, we again illustrate the prefactor $|g(\beta^\theta,\beta,r)\cos^{n}(2\theta)|^2$ whose inverse gives us the sample overhead in Fig.~\ref{fig:crosstalk}.
Although the crosstalk indeed increases the sample overhead compared to the ideal case, $\theta=0$, the effect is not significant as long as the crosstalk parameter $\theta$ is not sufficiently large.

\section{Fundamental limits for entanglement-free schemes}\label{app:efree}

\begin{figure}[!htb]
    \includegraphics[width=0.8\linewidth]{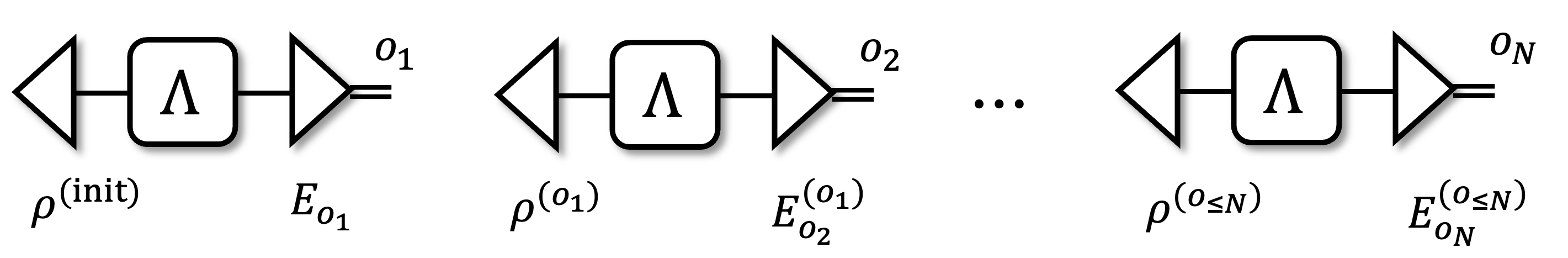}
    \caption{Schematics for entanglement-free schemes. In this work we assume no concatenation is allowed. Such a scheme can be completely specificed by a collection of input states and POVM measurements that adaptively depend on the measurement outcomes from the previous round.}
    \label{fig:scheme_SM}
\end{figure}

In this section, we prove the fundamental limit on general entanglement-free schemes for learning $n$-mode random displacement channels. 
In this work, we will focus on the ancilla-free schemes without concatenation. This means that, for each copy of the channel, we act it on some input state and apply a destructive POVM measurement right after. The input states and measurements can be adaptively chosen depending on previous measurement outcomes. See Fig.~\ref{fig:scheme_SM}. Bounds for such schemes have been investigated in different tasks~\cite{huang2021information,aharonov2022quantum,chen2021quantum}. One can also study ancilla-free protocols with concatenation, the lower bounds for which have been obtained in several recent works~\cite{chen2021quantum,chen2023tight,chen2023futility}, but we leave that for future study as continuous-variable system puts an additional level of complexity.
Throughout this work, we assume entanglement-free schemes to have no concatenation.

\begin{theorem}\label{th:main}
       Let $\Lambda$ be an arbitrary $n$-mode random displacement channel ($n\ge 8$) and consider an entanglement-free scheme that uses $N$ copies of $\Lambda$. After all measurements are completed, the scheme receives the query $\beta\in \mathbb{C}^n$ and returns an estimate $\tilde\lambda(\beta)$ of $\Lambda$'s characteristic function $\lambda(\beta)$. Suppose that, with success probability at least 2/3, $|\tilde{\lambda}(\beta)-\lambda(\beta)|\le \epsilon\le 0.24$ for all $\beta$ such that $|\beta|^2 \le n\kappa$. Then $N\ge 0.01\epsilon^{-2}(1+1.98\kappa)^{n}$.
\end{theorem}

Recall that an entanglement-assisted scheme can achieve the same task using $O(\epsilon^{-2})$ copies of channels given sufficient squeezing and $\kappa=O(1)$. Therefore, we establish an exponential separation between learning bosonic random displacement channels with and without entanglement.  In the following, we start proving this result in Sec.~\ref{app:warmup} and present a core lemma in Sec.~\ref{app:lemma}. We also prove a bound for learning with Gaussian schemes in Sec.~\ref{app:gaussian}, which might be of independent interest.

Before proceeding, let us specify some regularization conditions. We will only work with proper vectors (i.e., normalizable vector) in the Hilbert space and bounded operator acting on the Hilbert space.
That is to say, all the quantum states we considered can be expressed as a density operator $\hat\rho$ with trace $1$, and all the POVM element $\hat{E}$ is a bounded positive semi-definite operator satisfying $\hat E\le \hat I$.
Perhaps the most representative example that does not admit the above form is the perfect homodyne detection projector $|x\rangle\langle x|$, which represents projection onto the quadrature $x$. 
While $\ket x$ is not a proper vector in the relevant Hilbert space, it can be treated as a limit of proper vectors in any physical setting.
Concretely, homodyne detection is implemented by applying a 50:50 beam splitter between the input state and a strong local oscillator \cite{serafini2017quantum}, and the above improper projector $|x\rangle\langle x|$ is obtained by taking the limit where the power of the oscillator goes to infinity.
Therefore, in a reasonable physical setup, the actual projector is constructed with a proper vector in the Hilbert space, thus satisfying our assumptions.
We emphasize that our entanglement-assisted strategy also satisfies the same assumption as we regularize the Bell measurements with general-dyne detection with a parameter $s<\infty$, see Sec.~\ref{app:combine}.

\subsection{Lower bound for entanglement-free schemes}\label{app:warmup}


Given positive number $n\ge8$ and $\epsilon\leq 0.24$, we introduce a family of ``3-peak'' random displacement channels, defined by their characteristic functions,
\begin{align}\label{eq:lower_instance2}
    \Lambda_\gamma:~\lambda_{\gamma}(\beta)
    \coleq
    e^{-\frac{|\beta|^2}{2\sigma^2}}+2i\epsilon_0 e^{-\frac{|\beta-\gamma|^2}{2\sigma^2}}-2i\epsilon_0 e^{-\frac{|\beta+\gamma|^2}{2\sigma^2}},\quad \gamma\in\mathbb C^n,
\end{align}
where $\epsilon_0 \coleq \epsilon/0.98 \le0.25$. The corresponding distributions of displacements, computed via Fourier transformation, are
\begin{equation}
    \Lambda_\gamma:~p_\gamma(\alpha) = \left(\frac{2\sigma^2}{\pi}\right)^n e^{-2\sigma^2|\alpha|^2}\left(1+4\epsilon_0\sin(2(\Im[\gamma]\Re[\alpha]-\Re[\gamma]\Im[\alpha]))\right),
\end{equation}
from which we see that the typical strength of displacement is of order $1/\sigma$. Roughly, the smaller $\sigma$ is, the larger energy the channel carries.
We define $\Lambda_\text{dep}\coleq\Lambda_{0}$ as the CV analogy of the depolarizing channel, and the other $\Lambda_\gamma$ can be viewed as perturbed depolarizing channels.
The set of 3-peak channels with parameters $(\epsilon,\sigma)$ is denoted as $\bm\Lambda^{\epsilon,\sigma}_\mr{3\mhyphen peak}$.
With this, we are going to prove a strictly stronger result than Theorem~\ref{th:main}. That is, even if one knows the channel to be estimated is from the restricted family, $\bm\Lambda^{\epsilon,\sigma}_\mr{3\mhyphen peak}$, an exponential lower bound still applies.
\begin{theorem}\label{th:finite}
        Given positive numbers $n, \sigma, \kappa, \epsilon$ such that
    \begin{equation}\label{eq:thS2_cond}
        2\sigma^2 \le \max\left\{1-1.98\kappa,~0.99\kappa\left(\sqrt{1+{(0.99\kappa)^{-2}}}-1\right)\right\}, \quad n\ge8 , \quad \epsilon\le0.24.
    \end{equation}
    If there exists an entanglement-free scheme such that, after learning from $N$ copies of an $n$-mode random displacement channel $\Lambda\in\bm\Lambda^{\epsilon,\sigma}_\mr{3\mhyphen peak}$, and then receiving a query $\beta\in\mbb C^n$, can return an estimate $\tilde\lambda(\beta)$ of $\lambda(\beta)$ such that $|\tilde\lambda(\beta)-\lambda(\beta)|\le\epsilon$ with probability at least $2/3$ for all $\beta$ such that $|\beta|^2\le n\kappa$, then 
    \begin{equation}
        N\geq 0.01\epsilon^{-2} \left(1+\frac{1.98\kappa}{1+2\sigma^2}\right)^{n}.
    \end{equation}
\end{theorem}

\begin{figure*}[t]
	\centering
        \includegraphics[width=0.95\linewidth]{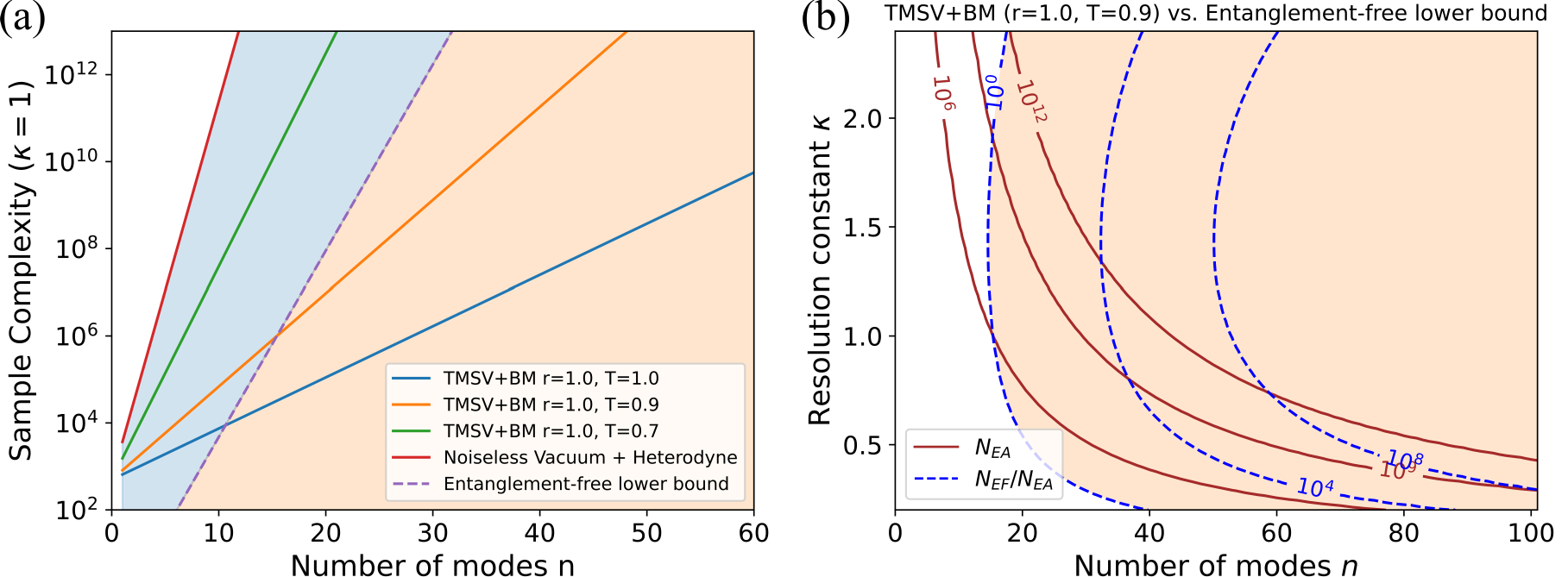}
        \caption{
        Learning random displacement channels from the family with $\sigma=0.3$ as in Theorem~\ref{th:finite}. (In the main text, we set $\sigma=0$.) All $\kappa$ shown in the figure satisfies Eq.~\eqref{eq:thS2_cond}.
        (a)~Comparison of TMSV+BM (with different loss rates), Vacuum+Heterodyne, and the entanglement-free lower bound at $\kappa=1$. The task is to estimate any $\lambda(\beta)$ such that $|\beta|^2\le\kappa n$ with precision $\varepsilon=0.2$ and success probability $1-\delta=2/3$. The orange region represents a rigorous advantage over any entanglement-free schemes. The blue region represents an advantage over Vacuum+heterodyne.
        (b)~Comparison of the TMSV+BM scheme with squeezing parameter $r=1.0$ and loss rate $1-T=0.1$ with the entanglement-free lower bound of Theorem~2. The task is the same as (a).
        The brown solid contour lines represent the sample complexity of TMSV+BM given by Theorem~3. The blue dashed contour lines represent the ratio of sample complexity between the entanglement-free lower bound and TMSV+BM, which clearly indicates the entanglement-enabled advantages.
	}
	\label{fig:result}
\end{figure*}

\noindent It is not hard to see that a $\sigma>0$ satisfying the assumptions can always be found for any $\kappa>0$. Indeed, Theorem~\ref{th:main} follows from Theorem~\ref{th:finite} by choosing $\sigma\to 0$. Note that Theorem~\ref{th:finite} does not place any constraint on the input state and measurement. This means that learning a finite-energy random displacement channel is hard without ancilla even given energy-unbounded input state and measurement. Also, Theorem~\ref{th:finite} enables an experimental test, as it only requires generating finite displacement with high probability. The practical performance of this bound with $\sigma=0.3$ is shown in Fig.~\ref{fig:result}.


\begin{proof}[Proof of Theorem~\ref{th:finite}]
Now we introduce the following game between Alice and Bob that helps reduce the learning task to a partially-revealed hypothesis testing task~\cite{huang2022quantum}. First, Alice samples $s\in\{\pm1\}$ with equal probability and $\gamma\in\mathbb C^n$ according to the multivariate normal distribution $q(\gamma)$ defined as
\begin{align}
    q(\gamma)\coleq \left(\frac{1}{2\pi \sigma_\gamma^2}\right)^n e^{-\frac{|\gamma|^2}{2\sigma_\gamma^2}},
\end{align}
where we will set $2\sigma_\gamma^2\coleq0.99{\kappa}$ to ensure the tail probability, i.e., $\Pr(|\gamma|^2>\kappa n)$, to be sufficiently small. Next, Alice does one of the following with equal probability:
\begin{enumerate}
    \item Prepare $N$ copies of $\Lambda_\text{dep}$ for Bob;
    \item Prepare $N$ copies of $\Lambda_{s\gamma}$ for Bob.
\end{enumerate}
Bob then measures the $N$ copies of the channels Alice prepared. 
After Bob has finished the measurements and retains only classical information, Alice reveals the value of $\gamma$ to Bob. 
Now Bob is asked to distinguish between the two hypotheses: whether Alice has prepared copies of $\Lambda_\text{dep}$ or $\Lambda_{s\gamma}$.
Crucially, Bob must have completed all quantum measurements before Alice reveals $\gamma$, and can only perform classical post-processing after that.



We first argue that if there is a scheme satisfying the assumptions of Theorem~\ref{th:finite}, then Bob can use it to win the game with an average probability much better than random guess. Bob's strategy is as follows: If the $\gamma$ he received satisfies $2\sigma^2<|\gamma|^2\le \kappa n$, use the scheme to query $\lambda(\gamma)$.
Note that for any $\gamma\in\mbb C^n$:
\begin{equation}
    \bigl|\lambda_\mr{dep}(\gamma) - \lambda_{\pm\gamma}(\gamma)\bigr| = \frac12\bigl|\lambda_{\gamma}(\gamma)-\lambda_{-\gamma}(\gamma)\bigr|  = 2\epsilon_0\left|1-e^{-\frac{4|\gamma|^2}{2\sigma^2}}\right|.
\end{equation}
For $|\gamma|^2>2\sigma^2$, the R.H.S. is lower bounded by $2\epsilon_0\times0.98 = 2\epsilon$. By assumption, this allows Bob to distinguish among $\{\Lambda_\mr{dep},\Lambda_\gamma,\Lambda_{-\gamma}\}$ and thus guess correctly with at least $2/3$ chance; For other $\gamma$, Bob just makes a uniformly random guess. Note that
\begin{align}
    \Pr\left(2\sigma^2<|\gamma|^2\le \kappa n\right) &= 1-\Pr(|\gamma|^2>\kappa n)-\Pr(|\gamma|^2\le2\sigma^2)\\
    &= 1- \left(\frac{\Gamma(n,n/0.99)}{\Gamma(n)}\right) - \left(1-\frac{\Gamma(n,\sigma^2/\sigma_\gamma^2)}{\Gamma(n)}\right)\\
    &\ge \frac12 - \left(1-\frac{\Gamma(n,\sigma^2/\sigma_\gamma^2)}{\Gamma(n)}\right)\label{eq:prob_gamma_inside}\\
    &=\frac12 - \frac{\int_0^{\sigma^2/\sigma^2_\gamma}t^{n-1}e^{-t}dt}{(n-1)!} \\
    &\ge 0.49987.
\end{align}
The first inequality is shown in Sec.~\ref{app:tail}.
The second inequality requires $n\ge8$ and $2\sigma^2\le0.99\kappa\coleq 2\sigma_\gamma^2$.
Bob's average success probability is lower bounded by 
\begin{equation}\label{eq:success_prob_lower}
  \Pr[\mr{Success}]\ge  \Pr\left(2\sigma^2<|\gamma|^2\le\kappa n\right)\times 2/3+\left(1-\Pr\left(2\sigma^2<|\gamma|^2\le\kappa n\right)\right)\times 1/2.
\end{equation}


\medskip

Now we investigate the probability distribution of Bob's measurement outcomes for any $\gamma$.
For any adaptive entanglement-free strategy, one specifies an input state and a POVM for the $i$th copy of $\Lambda$ that can depend on previous measurement outcomes. We denote the $i$th measurement outcomes as $o_i$ and the outcomes up to the $i$th round as $o_{<i} = [o_1,...,o_{i-1}]$. The latter is added as a superscript to the $i$th input states $\rho^{o_{<i}}$ and POVM element $E_{o_i}^{o_{<i}}$ to emphasize their adaptive nature. With these notations, the probability of obtaining outcomes $o_{1:N}$ on $N$ copies of $\Lambda$ is
\begin{equation}
    p(o_{1:N}|\Lambda) = \prod_{k=1}^N \Tr[{\hat E}_{o_i}^{o_{<i}}\Lambda({\hat\rho}^{o_{<i}})].
\end{equation}
For a fixed $\gamma$, let $p_1(o_{1:N}) \coleq p(o_{1:N}|\Lambda_\text{dep}),~ p_{2,\gamma}(o_{1:N}) \coleq \mathbb E_{s=\pm1}p(o_{1:N}|\Lambda_{s\gamma})$, which is the distribution of Bob's outcomes under the two hypotheses, respectively, conditioned on the $\gamma$ he received. According to the property of total variation distance, the maximal probability that Bob can distinguish $p_1$ and $p_{2,\gamma}$ is bounded by 
\begin{align}
    \Pr[\text{Success}|\gamma] \le \frac{1}{2}(1+\text{TVD}(p_1,p_{2,\gamma})),
\end{align}
where TVD is the \emph{total variation distance} defined as
\begin{align}
    \text{TVD}(p_1,p_{2,\gamma})
    \coleq \sum_{o_{1:N}}\max\left\{0,~p_1(o_{1:N})-p_{2,\gamma}(o_{1:N})\right\}.
\end{align}
We note that the sum over $o_{1:N}$ should be understood as integral for continuous-variable outcomes.

\medskip

\noindent Thus, the average probability that Bob can win the game is upper bounded by
\begin{equation}\label{eq:success_prob_upper}
    \Pr[\text{Success}] = \mathbb E_{\gamma\sim q}\Pr[\text{Success}|\gamma] \le \frac12(1+\mathbb E_\gamma \text{TVD}(p_1,p_{2,\gamma})).
\end{equation}
Combining Eq.~\eqref{eq:success_prob_lower} and Eq.~\eqref{eq:success_prob_upper}, we get
\begin{equation}\label{eq:TVD_lower}
    \mbb E_\gamma \mr{TVD}(p_1,p_{2,\gamma})\ge 0.1666.
\end{equation}
\noindent In the following, we show by direct calculation that this is impossible unless $N$ is exponentially large in $n$, which yields a desired lower bound for the sample complexity.

\medskip

Thanks to convexity, we assume pure input states and rank-1 measurement without decreasing the TVD, i.e., the $k$th round's input state and POVM are written as $|A^{o_{<k}}\rangle$ and $\{|B_{o_k}^{o_{<k}}\rangle\langle B_{o_k}^{o_{<k}}|\}$, which are conditioned on the previous measurement outcomes $o_{<k}$.
Here, the input state has unit length and $\sum_{o_k}|B_{o_k}^{o_{<k}}\rangle\langle B_{o_k}^{o_{<k}}|=\mathds{1}$.
We note that since any density matrix is trace-class, a spectrum decomposition always exists. 
On the other hand, the POVM element can be non-compact operator and might not have spectrum decomposition, but it is known that they can always be composed into rank-1 projectors with positive coefficients (see~\cite[Theorem~6]{kornelson2004rank}). Thus, making both the input state and measurement projector to be rank-1 is indeed justified.

\medskip

Now, let us rewrite the probabilities as
\begin{align}
    p_1(o_{1:N})&=\prod_{k=1}^N \langle B_{o_k}^{o_{<k}}|\Lambda_{\text{dep}}(|A^{o_{<k}}\rangle\langle A^{o_{<k}}|)|B_{o_k}^{o_{<k}} \rangle
    \\&=\prod_{k=1}^N\left(\frac{1}{\pi^n}\int d^{2n}\beta_k \lambda_{\text{dep}}(\beta_k)\langle B_{o_k}^{o_{<k}}|\hat{D}^\dagger(\beta_k)|B_{o_k}^{o_{<k}}\rangle\langle A^{o_{<k}}|\hat{D}(\beta_k)|A^{o_{<k}}\rangle\right), \\ 
    p_{2,\gamma}(o_{1:N})&= \mathbb E_{s={\pm 1}}\prod_{k=1}^N \langle B_{o_k}^{o_{<k}}|\Lambda_{s\gamma}(|A^{o_{<k}}\rangle\langle A^{o_{<k}}|)|B_{o_k}^{o_{<k}}\rangle
    \\&=\mathbb E_{s=\pm1}\prod_{k=1}^N\left(\frac{1}{\pi^{n}}\int d^{2n}\beta_k \lambda_{s\gamma}(\beta_k)\langle B_{o_k}^{o_{<k}}|\hat{D}^\dagger(\beta_k)|B_{o_k}^{o_{<k}}\rangle\langle A^{o_{<k}}|\hat{D}(\beta_k)|A^{o_{<k}}\rangle\right).
\end{align}
Let $\lambda_{\gamma}^{\text{add}}(\beta_k)\coleq\lambda_{\gamma}(\beta_k)- \lambda_{\text{dep}}(\beta_k)$.
The difference of the probabilities can then be written as
\begin{align}
    &p_1(o_{1:N})-p_{2,\gamma}(o_{1:N})\\
    =& p_1(o_{1:N})\left(1-\frac{p_{2,\gamma}(o_{1:N})}{p_1(o_{1:N})}\right) \\
    =&p_1(o_{1:N})\left(1-\mathbb{E}_{s=\pm1}\prod_{k=1}^N \left(1+\frac{\frac{1}{\pi^n}\int d^{2n}\beta_k \lambda_{s\gamma}^{\text{add}}(\beta_k)\langle B_{o_k}^{o_{<k}}|\hat{D}^\dagger(\beta_k)|B_{o_k}^{o_{<k}}\rangle\langle A^{o_{<k}}|\hat{D}(\beta_k)|A^{o_{<k}}\rangle}{\frac{1}{\pi^n}\int d^{2n}\beta_k \lambda_{\text{dep}}(\beta_k)\langle B_{o_k}^{o_{<k}}|\hat{D}^\dagger(\beta_k)|B_{o_k}^{o_{<k}}\rangle\langle A^{o_{<k}}|\hat{D}(\beta_k)|A^{o_{<k}}\rangle}\right)\right)  \\
    =& p_1(o_{1:N})\left(1-\mathbb{E}_{s=\pm1}\prod_{k=1}^N \left(1-4\epsilon_0\Im G_{\sigma}^{o_{\le k}}(s\gamma)\right)\right),\label{eq:diff_finite}
\end{align}
where we have defined
\begin{equation}\label{eq:def_G_sigma}
    G_\sigma^{o_{\le k}}(\gamma) \coleq \frac{\int d^{2n}\beta e^{-\frac{|\beta-\gamma|^2}{2\sigma^2}}G^{o_{\le k}}(\beta)}{\int d^{2n}\beta' e^{-\frac{|\beta'|^2}{2\sigma^2}}G^{o_{\le k}}(\beta')},
\end{equation}
where
\begin{align}\label{eq:def_G}
    G^{o_{\le k}}(\beta)\coleq \frac{\langle B_{o_k}^{o_{<k}}|\hat{D}^\dagger(\beta)|B_{o_k}^{o_{<k}}\rangle}{\langle B_{o_k}^{o_{<k}}|B_{o_k}^{o_{<k}}\rangle}\cdot\langle A^{o_{<k}}|\hat{D}(\beta)|A^{o_{<k}}\rangle,
\end{align}
which satisfies $G^{o_{\le k}}(\gamma) = G^{o_{\le k}}(-\gamma)^*$.

\medskip

\noindent We thus have 
\begin{equation}
    \mbb E_\gamma\mr{TVD}(p_1,p_{2,\gamma}) = \mbb E_\gamma\sum_{o_{1:N}}p_1(o_{1:N})\max\left\{0,~1-\mathbb{E}_{s=\pm1}\prod_{k=1}^N \left(1-4\epsilon_0\Im G_\sigma^{o_{\le k}}(s\gamma)\right)\right\}.
\end{equation}
\noindent Now we can lower bound the following term, 
\begin{align}
    &\mathbb E_{s=\pm 1}\prod_{k=1}^N\left(1-4\epsilon_0\Im G_\sigma^{o_{\le k}}(s\gamma)\right) \label{eq:adaptive09i} \\
    \geq~& \prod_{k=1}^N \sqrt{\left(1-4\epsilon_0\Im G_\sigma^{o_{\le k}}(+\gamma)\right) \left(1-4\epsilon_0\Im G_\sigma^{o_{\le k}}(-\gamma)\right)} \\
    =~& \prod_{k=1}^N \sqrt{1-16\epsilon_0^2\left(\Im G_\sigma^{o_{\le k}}(\gamma)\right)^2} \\
    \geq~& \prod_{k=1}^N \left(1-16\epsilon_0^2 \left(\mr{Im}G_\sigma^{o_{\le k}}(\gamma)\right)^ 2 \right) \\
    \geq~& 1- \sum_{k=1}^N 16\epsilon_0^2 |G_\sigma^{o_{\le k}}(\gamma)|^2,
\end{align}
where the second line uses the AM-GM inequality and the fact that the expression inside the bracket is the ratio of two conditional probabilities and is thus non-negative; the third line uses the fact that $\Im G_\sigma^{o_{\le k}}(\gamma)=-\Im G_\sigma^{o_{\le k}}(-\gamma)$; the fourth line uses $\sqrt{1 - x} \geq 1 - x, \forall \, 0 \leq x \leq 1$; and the final line uses the inequality $\prod_{i} (1 - x_i) \geq 1 - \sum_{i} x_i$ for all $0 \leq x_i \leq 1$.
Thus, we can get rid of the maximum in the expression of the average TVD as
\begin{align}
    \mbb E_\gamma\text{TVD}(p_1,p_{2,\gamma})\le\sum_{o_{1:N}}p_1(o_{1:N})\sum_{k=1}^N16\epsilon_0^2\mbb E_\gamma|G_\sigma^{o_{\le k}}(\gamma)|^2.\label{eq:non-con-tvd-finite}
\end{align}

\medskip

To further upper bound the R.H.S., we need the following Lemma~\ref{le:brilliant_lemma}. The lemma is analogous to Pauli twirling in discrete-variable systems but also takes finite energy into consideration. The proof of Lemma~\ref{le:brilliant_lemma} is given in Sec.~\ref{app:lemma}; Alternatively, when the input states and measurements are restricted to be Gaussian, a more straightforward calculation is possible, yielding different bounds, which we will present in Sec.~\ref{app:gaussian}.
\begin{lemma}\label{le:brilliant_lemma}
    For any $\ket{A^{o_{<k}}},\ket{B^{o_{<k}}_{o_k}}$ we have
    \begin{equation}
        \mathbb{E}_\gamma|G_\sigma^{o_{\le k}}(\gamma)|^2\leq \left(\frac{1+2\sigma^2}{1+2\sigma^2+4\sigma_\gamma^2}\right)^n,
    \end{equation}
    given that $\sigma^2 \le \max\left\{\frac12-2\sigma_\gamma^2,~\sigma_\gamma^2\left(\sqrt{1+\frac1{4\sigma_\gamma^4}}-1\right)\right\}$.
\end{lemma}
\noindent Thanks to Lemma~\ref{le:brilliant_lemma}, we get the following upper bound
\begin{equation}\label{eq:TVD_upper}
    \mbb E_\gamma\text{TVD}(p_1,p_{2,\gamma})\le\sum_{o_{1:N}}p_1(o_{1:N})\sum_{k=1}^N16\epsilon_0^2 
    \left(\frac{1+2\sigma^2}{1+2\sigma^2+4\sigma_\gamma^2}\right)^n
    =16N\epsilon_0^2 
    \left(\frac{1+2\sigma^2}{1+2\sigma^2+4\sigma_\gamma^2}\right)^n.
\end{equation}
Combining this with the lower bound in Eq.~\eqref{eq:TVD_lower} and substituting $\epsilon = 0.98\epsilon_0$,
\begin{equation}
    N\ge 0.01\epsilon^{-2}\left(1+\frac{4\sigma_\gamma^2}{1+2\sigma^2}\right)^n.
\end{equation}
By substituting $2\sigma_\gamma^2=0.99\kappa$, we obtain the lower bound as claimed in Theorem~\ref{th:finite}.
\end{proof}

In Fig.~\ref{fig:result}, we compare the upper bound of the TMSV+BM scheme to the derived lower bound of entanglement-free schemes.
In contrast to the main text, we set $\sigma=0.3$ to consider a more practical case for experimental realization in the near future.
To see how much energy is required to realize the 3-peak channel, one can easily check that
for a given $\sigma$, the corresponding single-mode depolarizing channel $\Lambda_0$ transforms a vacuum input state to a thermal state of mean photon number $1/2\sigma^2$.
Since this channel is a product channel, it implies that we need $1/2\sigma^2$ average photons per mode.
For our choice $\sigma=0.3$, $1/2\sigma^2\approx 5.56$.
Since the envelope determined by $\sigma$ has a larger contribution than $\gamma$ that determines the oscillation, we are required to produce approximately $1/2\sigma^2$ photon number on average.
It is worth emphasizing that for $\kappa\leq 2.5$, the choice satisfies the condition of Theorem~\ref{th:finite}.

\subsection{Proof of Lemma~\ref{le:brilliant_lemma}}\label{app:lemma}
In this section we prove Lemma~\ref{le:brilliant_lemma}. Let $\ket A,\ket B$ be arbitrary normalized pure states in the $n$-mode bosonic Hilbert space. Define
\begin{align}
        G(\beta) &\coleq \langle B| \hat D^\dagger(\beta) |B\rangle \langle A| \hat D(\beta) |A\rangle,\\
        G_\sigma(\gamma) &\coleq \frac{[\mc N_\sigma * G](\gamma)}{[\mc N_\sigma *G](0)} \coleq \frac{\int d^{2n}\beta\exp(-|\beta-\gamma|^2/2\sigma^2)G(\beta)}{\int d^{2n}\beta\exp(-|\beta|^2/2\sigma^2)G(\beta)}.
\end{align}
Here $*$ stands for convolution. We are going to prove the following inequality
\begin{align}
    \mathbb{E}_\gamma|G_\sigma(\gamma)|^2
    =\frac{\mathbb{E}_\gamma|[\mc N_\sigma * G](\gamma)|^2}{|[\mc N_\sigma *G](0)|^2}\leq \left(\frac{1+2\sigma^2}{1+2\sigma^2+4\sigma_\gamma^2}\right)^n,
\end{align}
where $\gamma \sim    q(\gamma)\coleq \left(\frac{1}{2\pi \sigma_\gamma^2}\right)^n \exp({-\frac{|\gamma|^2}{2\sigma_\gamma^2}})$ and $\sigma^2 \le \max\left\{\frac12-2\sigma_\gamma^2,~\sigma_\gamma^2\left(\sqrt{1+\frac1{4\sigma_\gamma^4}}-1\right)\right\}$. 

\medskip

\noindent First of all, write the following expression in the Fourier basis
\begin{align}
    [\mc N_\sigma * G](\gamma)
    =\frac1{\pi^{2n}}\int d^{2n}\omega e^{\omega^\dagger\gamma-\gamma^\dagger\omega}F_{[\mc N_\sigma * G]}(\omega)
    =\frac1{\pi^{2n}}\int d^{2n}\omega e^{\omega^\dagger\gamma-\gamma^\dagger\omega}F_{\mc N_\sigma}(\omega)F_{G}(\omega),
\end{align}
where the last equality uses the convolution theorem~\cite{castleman1996digital}. The Fourier component of $\mc N_\sigma$ is
\begin{align}
    F_{\mc N_\sigma}(\omega)&=\int d^{2n}\beta e^{-\frac{|\beta|^2}{2\sigma^2}}e^{\beta^\dagger \omega-\omega^\dagger \beta} = (2\pi\sigma^2)^n e^{-2\sigma^2|\omega|^2}.
\end{align}
The Fourier component of $G$ can be computed as
\begin{align}
    F_{G}(\omega)&=\int d^{2n}\beta \langle B| \hat D^\dagger(\beta) |B\rangle \langle A| \hat D(\beta) |A\rangle e^{\beta^\dagger\omega-\omega^\dagger\beta}\label{eq:_middle_step_2}\\
    &=\int d^{2n}\beta \langle B| \hat D^\dagger(\beta) |B\rangle \langle A| \hat D^\dagger(\omega)\hat D(\beta)\hat D(\omega) |A\rangle \\
    &=\pi^n|\langle B|\hat{D}(\omega)|A\rangle|^2,
\end{align}
where the second line uses $\hat D^\dagger(\omega)\hat D(\beta)\hat D(\omega)=e^{\beta^\dagger\omega-\omega^\dagger\beta}\hat D(\beta)$, and the last line is by Eq.~\eqref{eq:displacement_fourier}. 
Thus,
\begin{align}
    &|[\mc N_\sigma * G](\gamma)|^2 \\
    &=\left|\frac{1}{\pi^{2n}}\int d^{2n}\omega e^{\omega^\dagger\gamma-\gamma^\dagger\omega}F_{\mc N_\sigma}(\omega)F_{G}(\omega)\right|^2 \\ 
    &=\frac{1}{\pi^{4n}}\int d^{2n}\omega d^{2n}\omega' e^{(\omega-\omega')^\dagger\gamma-\gamma^\dagger(\omega-\omega')}F_{\mc N_\sigma}(\omega)F^*_{\mc N_\sigma}(\omega')F_{G}(\omega)F^*_{G}(\omega') \\ 
    &=(2\sigma^2)^{2n}\int d^{2n}\omega d^{2n}\omega' e^{(\omega-\omega')^\dagger\gamma-\gamma^\dagger(\omega-\omega')}e^{-2\sigma^2(|\omega|^2+|\omega'|^2)}|\langle B,B|\hat{D}(\omega)\otimes \hat{D}(\omega')|A,A\rangle|^2 \label{eq:_middle_step_1}.
\end{align}
After averaging over Gaussian distribution of $\gamma$, we obtain the numerator
\begin{align}
    \mathbb{E}_\gamma|[\mc N_\sigma * G](\gamma)|^2
    &=(2\sigma^2)^{2n}\int d^{2n}\omega d^{2n}\omega' e^{-2\sigma_\gamma^2|\omega-\omega'|^2}e^{-2\sigma^2(|\omega|^2+|\omega'|^2)}|\langle B,B|\hat{D}(\omega)\otimes \hat{D}(\omega')|A,A\rangle|^2 \\ 
    &=(2\sigma^2)^{2n}\int d^{2n}\alpha d^{2n}\beta e^{-4\sigma_\gamma^2|\beta|^2}e^{-2\sigma^2(|\alpha|^2+|\beta|^2)}|\langle B,B|\hat{U}_\text{BS}^\dagger\hat{D}(\alpha)\otimes \hat{D}(\beta)\hat{U}_\text{BS}|A,A\rangle|^2.
\end{align}
Here, we changed the variable as $\omega=(\alpha+\beta)/\sqrt{2}$ and $\omega'=(\alpha-\beta)/\sqrt{2}$, i.e., $(\omega+\omega')/\sqrt{2}=\alpha$ and $(\omega-\omega')/\sqrt{2}=\beta$ and chose the 50:50 beam splitter such that
\begin{align}
    \hat{U}_\text{BS}^\dagger \hat{D}(\alpha)\otimes \hat{D}(\beta)\hat{U}_\text{BS}=\hat{D}((\alpha+\beta)/\sqrt{2})\otimes \hat{D}((\alpha-\beta)/\sqrt{2}).
\end{align}
The denominator follows similarly from Eq.~\eqref{eq:_middle_step_1} as
\begin{align}
    |[\mc N_\sigma * G](0)|^2
    &=(2\sigma^2)^{2n}\int d^{2n}\omega d^{2n}\omega' e^{-2\sigma^2(|\omega|^2+|\omega'|^2)}|\langle B,B|\hat{D}(\omega)\otimes \hat{D}(\omega')|A,A\rangle|^2 \\
    &=(2\sigma^2)^{2n}\int d^{2n}\alpha d^{2n}\beta e^{-2\sigma^2(|\alpha|^2+|\beta|^2)}|\langle B,B|\hat{U}_\text{BS}^\dagger\hat{D}(\alpha)\otimes \hat{D}(\beta)\hat{U}_\text{BS}|A,A\rangle|^2.
\end{align}
To further simplify the expressions, note that by applying the convolution theorem to Eq.~\eqref{eq:_middle_step_2}, we have \begin{align}
    \frac{1}{\pi^n}|\langle B|\hat{D}(\alpha)|A\rangle|^2
    =\int d^{2n}\beta W_{A}(\beta)W_B(\beta-\alpha),
\end{align}
where $W_A$ and $W_B$ are the Wigner functions of the states $|A\rangle$ and $|B\rangle$, respectively.
\begin{figure*}[t]
	\centering
        \includegraphics[width=0.4\linewidth]{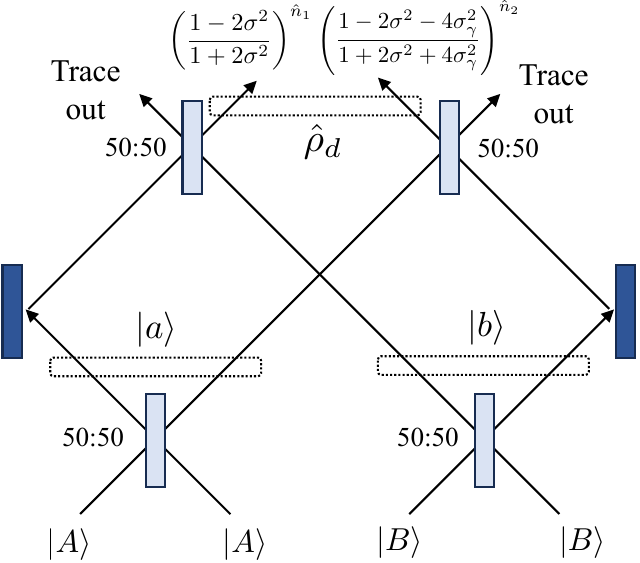}
        \caption{
        Schematics for Eq.~\eqref{eq:proof_i} to \eqref{eq:proof_f}.
        Here, each line represents $n$-mode state, and we omit the phase factors for simplicity.
	}
	\label{fig:proof}
\end{figure*}
Here, note the sign in the arguments due to the complex conjugate of the characteristic function, $\langle B|\hat{D}^\dagger(\beta)|B\rangle$, in Eq.~\eqref{eq:_middle_step_2}.
Thus, by defining the $2n$-mode states $|a\rangle\coleq \hat{U}_\text{BS}|A,A\rangle$ and $|b\rangle\coleq \hat{U}_\text{BS}|B,B\rangle$, we have
\begin{align}
    &\int d^{2n}\alpha d^{2n}\beta e^{-4\sigma_\gamma^2|\beta|^2}e^{-2\sigma^2(|\alpha|^2+|\beta|^2)}|\langle B,B|\hat{U}_\text{BS}^\dagger\hat{D}(\alpha)\otimes \hat{D}(\beta)\hat{U}_\text{BS}|A,A\rangle|^2 \label{eq:proof_i}\\ 
    &=\pi^{2n}\int d^{2n}\omega_1d^{2n}\omega_2 d^{2n}\alpha d^{2n}\beta e^{-2\sigma^2|\alpha|^2} e^{-(4\sigma_\gamma^2+2\sigma^2)|\beta|^2} W_a(\omega_1,\omega_2)W_{b}(\omega_1-\alpha,\omega_2-\beta) \\ 
    &=\pi^{2n}\int d^{2n}\omega_1d^{2n}\omega_2 d^{2n}\gamma_1 d^{2n}\gamma_2 e^{-2\sigma^2|\omega_1-\gamma_1|^2} e^{-(4\sigma_\gamma^2+2\sigma^2)|\omega_2-\gamma_2|^2} W_a(\omega_1,\omega_2)W_{b}(\gamma_1,\gamma_2) \\ 
    &=\pi^{2n}\int d^{2n}\alpha_1d^{2n}\alpha_2 d^{2n}\beta_1 d^{2n}\beta_2 e^{-4\sigma^2|\alpha_1|^2} e^{-(8\sigma_\gamma^2+4\sigma^2)|\alpha_2|^2} W_a\left(\frac{\alpha_1+\beta_1}{\sqrt{2}},\frac{\alpha_2+\beta_2}{\sqrt{2}}\right)W_{b}\left(\frac{\beta_1-\alpha_1}{\sqrt{2}},\frac{\beta_2-\alpha_2}{\sqrt{2}}\right) \\ 
    &=\pi^{2n}\int d^{2n}\alpha_1 d^{2n}\alpha_2  e^{-4\sigma^2|\alpha_1|^2} e^{-(8\sigma_\gamma^2+4\sigma^2)|\alpha_2|^2} W_d(\alpha_1,\alpha_2) \\
    &=\pi^{2n}(1+2\sigma^2)^{-n}(1+2\sigma^2+4\sigma^2_\gamma)^{-n}\Tr[\hat{\rho}_d\left(\frac{1-2\sigma^2}{1+2\sigma^2}\right)^{\hat{n}_1}\otimes \left(\frac{1-2\sigma^2-4\sigma^2_\gamma}{1+2\sigma^2+4\sigma^2_\gamma}\right)^{\hat{n}_2}], \label{eq:proof_f}
\end{align}
where we used the convolution theorem for the first equality, and we changed the variables for the second and third equalities, and
\begin{align}
    W_d(\alpha_1,\alpha_2)
    =\int d^{2n}\beta_1 d^{2n}\beta_2 W_a\left(\frac{\alpha_1+\beta_1}{\sqrt{2}},\frac{\alpha_2+\beta_2}{\sqrt{2}}\right)W_{b}\left(\frac{\beta_1-\alpha_1}{\sqrt{2}},\frac{\beta_2-\alpha_2}{\sqrt{2}}\right)
\end{align}
is the Wigner function of the state $\hat{\rho}_d$ obtained by applying a 50:50 beam splitter to the state $|a\rangle$ and $|b\rangle$ and tracing out half of the output.
For the last equality, $\hat{n}_1$ and $\hat{n}_2$ are the sum of the photon number operators for the first and second $n$ modes, respectively, and we use the following correspondence between the Wigner function and the operator
\begin{equation}
    \frac{e^{-4x|\alpha|^2}}{\pi^n}
    \Longleftrightarrow  
    \frac{\left[(1-2x)/(1+2x)\right]^{\hat{n}}}{(1+2x)^n}, 
\end{equation}
for any $x>0$ (see e.g.~\cite[Eq.~(3.6.39)]{barnett2002methods}). Note that, when $x>1/2$ the R.H.S. is proportional to a thermal state. Similar methods have been used to prove the maximum fidelity of Gaussian channels~\cite{caves2004fidelity}.
We illustrate the procedure in Fig.~\ref{fig:proof}.
With the same logic, we have
\begin{align}
    &\int d^{2n}\alpha d^{2n}\beta e^{-2\sigma^2(|\alpha|^2+|\beta|^2)}|\langle B,B|\hat{U}_\text{BS}^\dagger\hat{D}(\alpha)\otimes \hat{D}(\beta)\hat{U}_\text{BS}|A,A\rangle|^2 \\
    &=\pi^{2n}(1+2\sigma^2)^{-2n}\Tr\left[\hat{\rho}_d\left(\frac{1-2\sigma^2}{1+2\sigma^2}\right)^{\hat{n}_1}\otimes \left(\frac{1-2\sigma^2}{1+2\sigma^2}\right)^{\hat{n}_2}\right].
\end{align}
Hence, we have
\begin{align}
    \mathbb{E}_\gamma|G_\sigma(\gamma)|^2
    =\left(\frac{1+2\sigma^2}{1+2\sigma^2+4\sigma^2_\gamma}\right)^n\frac{\Tr\left[\hat{\rho}_d\left(\frac{1-2\sigma^2}{1+2\sigma^2}\right)^{\hat{n}_1}\otimes \left(\frac{1-2\sigma^2-4\sigma^2_\gamma}{1+2\sigma^2+4\sigma^2_\gamma}\right)^{\hat{n}_2}\right]}{\Tr\left[\hat{\rho}_d\left(\frac{1-2\sigma^2}{1+2\sigma^2}\right)^{\hat{n}_1}\otimes \left(\frac{1-2\sigma^2}{1+2\sigma^2}\right)^{\hat{n}_2}\right]}.
\end{align}
We now consider two parameter regimes. First, if  $2\sigma^2+4\sigma_\gamma^2\leq 1$, the operators on the R.H.S. are positive-semidefinite, and it is not hard to see, by monotonicity, that 
\begin{align}
    \frac{\Tr\left[\hat{\rho}_d\left(\frac{1-2\sigma^2}{1+2\sigma^2}\right)^{\hat{n}_1}\otimes \left(\frac{1-2\sigma^2-4\sigma^2_\gamma}{1+2\sigma^2+4\sigma^2_\gamma}\right)^{\hat{n}_2}\right]}{\Tr\left[\hat{\rho}_d\left(\frac{1-2\sigma^2}{1+2\sigma^2}\right)^{\hat{n}_1}\otimes \left(\frac{1-2\sigma^2}{1+2\sigma^2}\right)^{\hat{n}_2}\right]}\leq 1.
\end{align}
Second, if $2\sigma^2+4\sigma_\gamma^2>1$ but 
$2\sigma^2\le2\sigma_\gamma^2\left(\sqrt{1+\frac1{4\sigma_\gamma^4}}-1\right)\le1$ (the last inequality holds for all $\sigma_\gamma>0$), the above can be bounded as
\begin{align}
    \frac{\Tr\left[\hat{\rho}_d\left(\frac{1-2\sigma^2}{1+2\sigma^2}\right)^{\hat{n}_1}\otimes \left(\frac{1-2\sigma^2-4\sigma^2_\gamma}{1+2\sigma^2+4\sigma^2_\gamma}\right)^{\hat{n}_2}\right]}{\Tr\left[\hat{\rho}_d\left(\frac{1-2\sigma^2}{1+2\sigma^2}\right)^{\hat{n}_1}\otimes \left(\frac{1-2\sigma^2}{1+2\sigma^2}\right)^{\hat{n}_2}\right]}
    &\leq 
    \frac{\Tr\left[\hat{\rho}_d\left(\frac{1-2\sigma^2}{1+2\sigma^2}\right)^{\hat{n}_1}\otimes \left|\frac{1-2\sigma^2-4\sigma^2_\gamma}{1+2\sigma^2+4\sigma^2_\gamma}\right|^{\hat{n}_2}\right]}{\Tr\left[\hat{\rho}_d\left(\frac{1-2\sigma^2}{1+2\sigma^2}\right)^{\hat{n}_1}\otimes \left(\frac{1-2\sigma^2}{1+2\sigma^2}\right)^{\hat{n}_2}\right]} \\ 
    &=\frac{\Tr\left[\hat{\rho}_d\left(\frac{1-2\sigma^2}{1+2\sigma^2}\right)^{\hat{n}_1}\otimes \left(\frac{1-2\Sigma^2}{1+2\Sigma^2}\right)^{\hat{n}_2}\right]}{\Tr\left[\hat{\rho}_d\left(\frac{1-2\sigma^2}{1+2\sigma^2}\right)^{\hat{n}_1}\otimes \left(\frac{1-2\sigma^2}{1+2\sigma^2}\right)^{\hat{n}_2}\right]} \\
    &\leq 1,
\end{align}
where, in the second line, we define $-\frac{1-2\sigma^2-4\sigma_\gamma^2}{1+2\sigma^2+4\sigma_\gamma^2}\coleq\frac{1-2\Sigma^2}{1+2\Sigma^2}$, i.e., $\Sigma^2=\frac{1}{4(\sigma^2+2\sigma_\gamma^2)}$.
In the third line, we use $\Sigma^2\ge\sigma^2$, which can be easily verified under our assumptions for $\sigma$.
Therefore, as long as $\sigma^2 \le \max\left\{\frac12-2\sigma_\gamma^2,~\sigma_\gamma^2\left(\sqrt{1+\frac1{4\sigma_\gamma^4}}-1\right)\right\}$, we have the following bound
\begin{align}
    \mathbb{E}_\gamma|G_\sigma(\gamma)|^2\leq \left(\frac{1+2\sigma^2}{1+2\sigma^2+4\sigma^2_\gamma}\right)^n.
\end{align}
This completes the proof of Lemma~\ref{le:brilliant_lemma}. Note that the equality can be achieved if $\hat{\rho}_d$ is chosen to be the vacuum state. 
One can verify this holds when $\ket A=\ket B=\ket \alpha$ for some coherent state $\ket\alpha$.

\subsection{Lower bound for entanglement-free Gaussian schemes}\label{app:gaussian}



In this section, we give a lower bound for a specific class of scheme, the \emph{Gaussian schemes}, which may be of independent interest.
An ancilla-free Gaussian scheme is specified by collections of adaptively chosen Gaussian input state and Gaussian measurements. Again, thanks to convexity, we again consider only pure input states and rank-$1$ POVM measurements.
A Gaussian input state can be expressed as $|A\rangle=\hat{D}(\omega)|\bar{A}\rangle$, where $|\bar{A}\rangle$ is a centered (\textit{i.e.}, zero-mean) Gaussian state; A Gaussian POVM can be written as
\begin{align}
    \hat{\Pi}(\alpha)
    =|B\rangle\langle B|
    =\frac{1}{\pi^{n}}\hat{D}(\alpha)|\bar{B}\rangle\langle \bar{B}|\hat{D}^\dagger(\alpha),
\end{align}
for outcomes $\alpha\in\mathbb{C}^{n}$, where $|\bar{B}\rangle$ is a centered Gaussian state. We refer the readers to Ref.~\cite{ferraro2005gaussian, weedbrook2012gaussian, serafini2017quantum} for more details about Gaussian quantum information.

\begin{proposition}\label{prop:finite}
    Given positive numbers $n, \sigma, \kappa, \epsilon$ such that
    \begin{equation}
        n\ge8 , \quad \epsilon\le0.24.
    \end{equation}
    If there exists an entanglement-free Gaussian scheme such that, after learning from $N$ copies of an $n$-mode random displacement channel $\Lambda\in\bm\Lambda^{\epsilon,\sigma}_\mr{3\mhyphen peak}$, and then receiving a query $\beta\in\mbb C^n$, can return an estimate $\tilde\lambda(\beta)$ of $\lambda(\beta)$ such that $|\tilde\lambda(\beta)-\lambda(\beta)|\le\epsilon$ with probability at least $2/3$ for all $\beta$ such that $|\beta|^2\le n\kappa$, then 
    \begin{equation}
        N\geq 0.01\epsilon^{-2}\min\left\{\left(1+\frac{0.99\kappa}{\sigma^2}\right)^{n/2},~\left(1+\frac{1.98\kappa}{1+2\sigma^2}\right)^{n}\right\}.
    \end{equation}
\end{proposition}

\noindent A few remarks before presenting the proof: When $\kappa=O(1)$ and $\sigma^2\ll \kappa$, the second expression in the minimization dominates, and we recover Theorem~\ref{th:finite}.
On the other hand, the bound for Gaussian schemes also holds for arbitrarily large $\sigma$, though the upper bound will enter a different branch and the separation with entanglement-assisted schemes becomes weaker.

\medskip

\begin{proof}


Consider the same partially-revealed hypothesis-testing task and the same strategy used by Bob in the proof of Theorem~\ref{th:finite}. 
Recall that the average TVD under the two hypotheses is lower bounded by
\begin{equation}\label{eq:gaussian_lower_tvd}
\begin{aligned}    
\mbb E_\gamma\mr{TVD}(p_1,p_{2,\gamma}) 
&\ge 0.1666.
\end{aligned}
\end{equation}
To upper bound the average TVD, recall the following bound derived in Eq.~\eqref{eq:non-con-tvd-finite},
\begin{align}
    \mbb E_\gamma\text{TVD}(p_1,p_{2,\gamma})\le\sum_{o_{1:N}}p_1(o_{1:N})\sum_{k=1}^N16\epsilon_0^2\mbb E_\gamma|G_\sigma^{o_{\le k}}(\gamma)|^2,\label{eq:non-con-tvd-finite-2}
\end{align}
with $G_\sigma^{o_{\le k}}$ defined as
\begin{align}
    G_\sigma^{o_{\le k}}(\gamma) &= \frac{\int d^{2n}\beta e^{-\frac{|\beta-\gamma|^2}{2\sigma^2}}G^{o_{\le k}}(\beta)}{\int d^{2n}\beta' e^{-\frac{|\beta'|^2}{2\sigma^2}}G^{o_{\le k}}(\beta')},\\
    G^{o_{\le k}}(\beta)&= \frac{\langle B_{o_k}^{o_{<k}}|\hat{D}^\dagger(\beta)|B_{o_k}^{o_{<k}}\rangle}{\langle B_{o_k}^{o_{<k}}|B_{o_k}^{o_{<k}}\rangle}\cdot\langle A^{o_{<k}}|\hat{D}(\beta)|A^{o_{<k}}\rangle,
\end{align}
Note that this bound holds for any $\sigma$. Now we calculate the R.H.S. with Gaussian schemes. First compute $G^{o_{\leq k}}(\beta)$,
\begin{align}
    G^{o_{\leq k}}(\beta)
    =\langle \bar{B}|\hat{D}^\dagger(\alpha)\hat{D}^\dagger(\beta)\hat{D}(\alpha)|\bar{B}\rangle\langle A|\hat{D}(\beta)|A\rangle
    =\langle \bar{B}|\hat{D}^\dagger(\beta)|\bar{B}\rangle\langle \bar{A}|\hat{D}(\beta)|\bar{A}\rangle e^{\beta^\dagger(\alpha-\omega)-(\alpha-\omega)^\dagger\beta},
\end{align}
Here, without loss of generality, we can always write $|\bar{A}\rangle=\hat{U}_{\text{BS}_A}\hat{U}_{\text{sq}_A}|0\rangle$, where $\hat{U}_{\text{BS}_A}$ represents the unitary operator for a beam-splitter network and $\hat{U}_{\text{sq}_A}$ represents the product of single-mode squeezing operations.
Similarly, $|\bar{B}\rangle=\hat{U}_{\text{BS}_B}\hat{U}_{\text{sq}_B}\ket 0$.

To simplify $G^{o_{\leq k}}(\beta)$, using $\hat a \coleq (\hat x+i\hat p)/\sqrt2$, we can rewrite the displacement operator as
\begin{align}
    \hat{D}(\beta)\coleq\exp(\beta\hat{a}^\dagger-\beta^\dagger\hat{a})
    =\exp(\sqrt{2}i\Im\beta \hat{x}-\sqrt{2}i\Re\beta\hat{p})
    =\exp(\sqrt{2}iv\cdot\hat{q}),
\end{align}
where $\hat{q}\coleq(\hat{x}_1,\dots,\hat{x}_n,\hat{p}_1,\dots,\hat{p}_n)^\text{T}$ and $v(\beta)\coleq (\Im\beta_1,\dots,\Im\beta_n,\Re\beta_1,\dots,\Re\beta_n)^\text{T}$.
And
\begin{align}
    \langle 0|\hat{D}(\beta)|0\rangle
    =e^{-\frac{1}{2}|\beta|^2}
    =e^{-\frac{1}{2}|v|^2}.
\end{align}
Now, let us introduce the symplectic matrix $S$ that describes the dynamics of quadrature operators under Gaussian unitary operation $\hat{U}$:
\begin{align}
    \hat{U}^\dagger\hat{q}_i\hat{U}=(S\hat{q})_i.
\end{align}
Since the Gaussian unitary operation we consider is written as $\hat{U}_\text{BS}\hat{U}_{\text{sq}}$, the symplectic matrix can be decomposed as $S=S_{\text{BS}}S_{\text{sq}}$.
Here, $S_{\text{BS}}$ is an orthogonal matrix and $\hat{S}_{\text{sq}}$ can be explicitly written as $\text{diag}(e^{r_1},\dots,e^{r_n},e^{-r_1},\dots,e^{-r_n})$, where $r_1,\dots,r_n\geq 0$ represent squeezing parameters for each mode.
We use $r$ for squeezing parameters for $|A\rangle$ and $s$ for $|\bar{B}\rangle$.
After the symplectic transformation $S$, the displacement operator transforms as
\begin{align}
    \exp(\sqrt{2}iv^\text{T}\hat{q})
    \to\exp(\sqrt{2}iv\cdot(S\hat{q}))
    =\exp(\sqrt{2}i(S^\text{T}v)\cdot\hat{q}),
\end{align}
and
\begin{align}
    \langle 0|\exp(\sqrt{2}i(S^\text{T}v)\cdot\hat{q})|0\rangle
    =e^{-\frac{1}{2}|S^\text{T}v|^2}
\end{align}
Thus,
\begin{align}
    \langle \bar{A}|\hat{D}(\beta)|\bar{A}\rangle
    =\langle 0|\hat{U}^\dagger_{\text{sq}_A}\hat{U}^\dagger_{\text{BS}_A}\hat{D}(\beta)\hat{U}_{\text{BS}_A}\hat{U}_{\text{sq}_A}|0\rangle
    =e^{-\frac{1}{2}|S_A^\text{T}v|^2},
\end{align}
and
\begin{align}
    \langle \bar{B}|\hat{D}^\dagger(\beta)|\bar{B}\rangle
    =\langle \bar{B}|\hat{D}(\beta)|\bar{B}\rangle
    =e^{-\frac{1}{2}|S_B^\text{T}v'|^2}
    =e^{-\frac{1}{2}|S_B^\text{T}v|^2},
\end{align}
where $v\coleq v(\beta)$.
The first equality is due to the fact that $\langle \bar{B}|\hat{D}^\dagger(\beta)|\bar{B}\rangle$ is real.
And we can write the phase factor as
\begin{align}
    \exp\left(\beta^\dagger (\alpha-\omega)-\beta(\alpha-\omega)^\dagger\right)
    =\exp(2iv^\text{T}u),
\end{align}
where $u\coleq(-\Re(\alpha_1-\omega_1),\dots,-\Re(\alpha_n-\omega_n),\Im(\alpha_1-\omega_1),\dots,\Im(\alpha_n-\omega_n))^\text{T}$.
Thus, 
\begin{align}
    G^{o_{\leq k}}(\beta)
    &=\langle \bar{B}|\hat{D}^\dagger(\beta)|\bar{B}\rangle\langle \bar{A}|\hat{D}(\beta)|\bar{A}\rangle e^{\beta^\dagger(\alpha-\omega)-(\alpha-\omega)^\dagger\beta} \\
    &=\exp\left[-\frac{1}{2}v^\text{T}(S_AS_A^\text{T}+S_BS_B^\text{T})v+2iv^\text{T} u\right] \\
    &\coleq\exp\left[-\frac{1}{2}v^\text{T}\Sigma v+2iv^\text{T} u\right] \\
    &=\exp\left[-\frac{1}{2}(Ov)^\text{T}D(Ov)+2i(Ov)^\text{T}\cdot (Ou)\right] \\
    &=\prod_{i=1}^{2n}\exp\left[-\frac{d_i}{2}\left(v_i'-\frac{2i}{d_i}u_i'\right)^2-\frac{2u_i'^{2}}{d_i}\right],
\end{align}
where $\Sigma\coleq S_AS_A^\text{T}+S_BS_B^\text{T}>0$ is diagonalized as $\Sigma=O^\text{T}DO$ with diagonal matrix $D=\text{diag}(d_1,\dots,d_{2n})> 0$, and $v'\coleq Ov$, $u'\coleq Ou$.
Let us analyze the spectrum of $D$.
Note that since $S_AS_A^\text{T}$ and $S_BS_B^\text{T}$ are physical covariance matrices, $S_AS_A^\text{T}\geq i\Omega$, $S_BS_B^\text{T}\geq i\Omega$, and thus $\Sigma\geq 2i\Omega$~\cite{serafini2017quantum}, where
\begin{align}
    \Omega=
    \begin{pmatrix}
        0 & 1 \\ 
        -1 & 0
    \end{pmatrix}
    \otimes \mathds{1}_M.
\end{align}
Hence, by the Williamson decomposition~\cite{serafini2017quantum} and Weyl's monotonicity principle~\cite{zhan2013matrix}, the spectrum of $\Sigma$ is composed of pairs such that the product of the $i$th and $(i+n)$th eigenvalues of this matrix is no smaller than $4$.
Without loss of generality, we label the eigenvalues $d_1,...,d_{2n}$ in such a way that $d_id_{i+n}\ge4$ for all $1\le i\le n$.



\medskip

\noindent Now let us compute $G_\sigma^{o_{\le k}}(\gamma)$.
\begin{align}
    G_\sigma^{o_{\le k}}(\gamma) = \frac{\int d^{2n}\beta e^{-\frac{|\beta-\gamma|^2}{2\sigma^2}}G^{o_{\le k}}(\beta)/(2\pi\sigma^2)^n}{\int d^{2n}\beta' e^{-\frac{|\beta'|^2}{2\sigma^2}}G^{o_{\le k}}(\beta')/(2\pi\sigma^2)^n}
\end{align}
By defining $z\coleq (\Im \gamma_1,\dots,\Im\gamma_n,\Re \gamma_1,\dots,\Re\gamma_n)$, and $z'\coleq Oz$, we have
\begin{align}
    &\frac{1}{(2\pi\sigma^2)^n}\int d^{2n}\beta e^{-\frac{|\beta-\gamma|^2}{2\sigma^2}}G^{o_{\le k}}(\beta)\\
    =~&\frac{1}{(2\pi\sigma^2)^n}\int d^{2n}v' \prod_{i=1}^{2n}\exp\left[-\frac{(v_i'-z_i')^2}{2\sigma^2}-\frac{d_i}{2}\left(v_i'-\frac{2i}{d_i}u_i'\right)^2-\frac{2u_i'^{2}}{d_i}\right] \\
    =~&\prod_{i=1}^{2n}\left[\frac{1}{\sqrt{1+d_i\sigma^2}}\exp(-\frac{2u_i'^2\sigma^2}{1+d_i\sigma^2})\exp\left(\frac{1}{2}\left(\frac{-d_iz_i'^2+4iu_i z'_i}{1+d_i\sigma^2}\right)\right)\right],
\end{align}
\noindent Therefore, we obtain
\begin{align}
    G_\sigma^{o_{\le k}}(\gamma)
    =\prod_{i=1}^{2n}\exp\left[\frac{1}{2}\left(\frac{-d_iz_i'^2+4iu_i z'_i}{1+d_i\sigma^2}\right)\right],~~~\text{and}~~~
    |G_\sigma^{o_{\le k}}(\gamma)|
    =\prod_{i=1}^{2n}\exp\left[\frac{-d_iz_i'^2}{2(1+d_i\sigma^2)}\right].
\end{align}
Thus, after taking the average over $\gamma$, we obtain
\begin{align}
    \mathbb{E}_\gamma|G_\sigma^{o_{\le k}}(\gamma)|^2
    =\int d^{2n}z' \frac{e^{-\frac{|z|^2}{2\sigma_\gamma^2}}}{(2\pi\sigma_\gamma^2)^n}\prod_{i=1}^{2n}\exp\left[\frac{-d_iz_i'^2}{(1+d_i\sigma^2)}\right]
    =\prod_{i=1}^{2n}\sqrt{\frac{1}{1+2d_i\sigma_\gamma^2/(1+d_i\sigma^2))}}.
\end{align}
Substituting this back to Eq.~\eqref{eq:non-con-tvd-finite-2}, we obtain the following upper bound for the average TVD
\begin{equation}
    \mbb E_\gamma \mr{TVD}(p_1,p_{2,\gamma}) \le 16N\epsilon_0^2\prod_{i=1}^{2n}\sqrt{\frac{1}{1+2d_i\sigma_\gamma^2/(1+d_i\sigma^2))}},
\end{equation}
which, combined with the lower bound Eq.~\eqref{eq:gaussian_lower_tvd} and substituting $\epsilon_0=\epsilon/0.98$, yields the following sample complexity bound
\begin{equation}
    N \ge 0.01\epsilon^{-2}\left(\prod_{i=1}^{2n}\sqrt{1+\frac{2d_i\sigma_\gamma^2}{1+d_i\sigma^2}}\right). 
\end{equation}
To find a lower bound independent of $d_i$'s, focus on the following product, for any $1\le i\le n$,
\begin{align}
    \left(1+\frac{2d_i\sigma_\gamma^2}{1+d_i\sigma^2}\right)
    \left(1+\frac{2d_{i+n}\sigma_\gamma^2}{1+d_{i+n}\sigma^2}\right)
\end{align}
This is an increasing function in $d_i$ and $d_{i+n}$. 
We know the spectrum satisfies $d_i d_{i+n}\geq 4$.
Hence, we can lower bound it by setting $d_{i+n}/2=2/d_i\coleq d>0$, which leads to
\begin{align}
    \left(1+\frac{2d_i\sigma_\gamma^2}{1+d_i\sigma^2}\right)
    \left(1+\frac{2d_{i+n}\sigma_\gamma^2}{1+d_{i+n}\sigma^2}\right)\geq \frac{(d+2\sigma^2+4\sigma^2_\gamma)(1+2d(\sigma^2+2\sigma^2_\gamma))}{(d+2\sigma^2)(1+2d\sigma^2)}.
\end{align}
The R.H.S. is differentiable in $d$, with its only extreme value at $d=1$ being
\begin{align}
    \left(1+\frac{4\sigma_\gamma^2}{1+2\sigma^2}\right)^2.
\end{align}
Meanwhile, when $d\to0$ or $d\to\infty $, it becomes $1+2\sigma_\gamma^2/\sigma^2$.
We thus have the following lower bound,
\begin{align}
    \left(1+\frac{2d_i\sigma_\gamma^2}{1+d_i\sigma^2}\right)
    \left(1+\frac{2d_{i+n}\sigma_\gamma^2}{1+d_{i+n}\sigma^2}\right)\geq \min\left\{1+\frac{2\sigma_\gamma^2}{\sigma^2}, \left(1+\frac{4\sigma_\gamma^2}{1+2\sigma^2}\right)^2\right\},
\end{align}
which gives us the following sample complexity lower bound:
\begin{align}
    N\geq 0.01\epsilon^{-2}\min\left\{\left(1+\frac{2\sigma_\gamma^2}{\sigma^2}\right)^{n/2},~\left(1+\frac{4\sigma_\gamma^2}{1+2\sigma^2}\right)^{n}\right\}. 
\end{align}
Substituting $2\sigma_\gamma^2=0.99\kappa$ completes the proof of Proposition~\ref{prop:finite}.

\end{proof}

\section{Gaussian tail effect} \label{app:tail}

\begin{figure}[!htb]
    \includegraphics[width=0.45\linewidth]{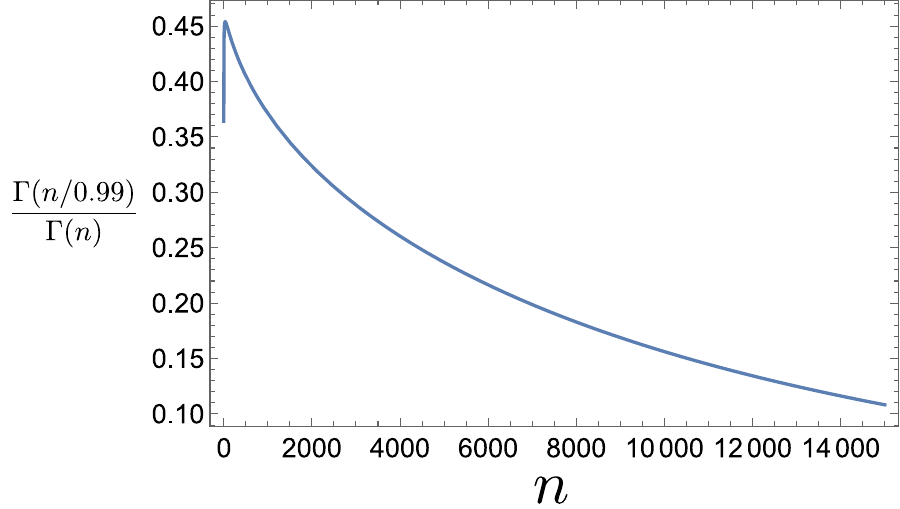}
    \caption{Numerical verification that the Gaussian tail probability is upper bounded by 0.5 for $n$ up to 14000.}
    \label{fig:tail}
\end{figure}

In this section, we find the condition that the effect of truncating a multivariate normal distribution is smaller than 0.5, which is used to derive the lower bound for entanglement-free schemes.
Consider a multivariate normal distribution:
\begin{align}
    q(\bm{x})=\left(\frac{1}{2\pi\sigma^2}\right)^n\exp\left(-\frac{|\bm{x}|^2}{2\sigma^2}\right),
\end{align}
where $\bm{x}\in \mathbb{R}^{2n}$.
Note that in the main text, while we consider $\mathbb{\gamma}\in \mathbb{C}^n$, they are equivalent.
Now, we consider a truncated distribution with $|\bm{x}|^2\leq R^2$ with a given $R$:
\begin{align}\label{eq:gaussian_tail_gamma}
    \int_{|\bm{x}|\leq R} d\bm{x} q(\bm{x})
    &=\int_{|\bm{x}|\leq R} d\bm{x} \left(\frac{1}{2\pi\sigma^2}\right)^n\exp\left(-\frac{|\bm{x}|^2}{2\sigma^2}\right) \\ 
    &=\left(\frac{1}{2\pi\sigma^2}\right)^n\int_{0}^R dr \int d\Omega_{2n} r^{2n-1}\exp\left(-\frac{r^2}{2\sigma^2}\right) \\ 
    &=1-\frac{\Gamma\left(n,\frac{R^2}{2\sigma^2}\right)}{\Gamma(n)}.
\end{align}
where we have used the following integrals:
\begin{align}
    \int d\Omega_{n}=\frac{2\pi^{n/2}}{\Gamma(n/2)},~~~~~
    \int_{0}^R dr r^{2n-1}\exp\left(-\frac{r^2}{2\sigma^2}\right)
    =2^{n-1}\sigma^{2n}\left[\Gamma(n)-\Gamma\left(n,\frac{R^2}{2\sigma^2}\right)\right],
\end{align}
and
\begin{align}
    \Gamma(n,x)=\int_x^\infty t^{n-1}e^{-t}dt
\end{align}
is the (upper) incomplete gamma function and $\Gamma(n)=\Gamma(n,0)$.
Therefore, the tail probability is given by $\Gamma\left(n,\frac{R^2}{2\sigma^2}\right)/{\Gamma(n)}$.
In the main text and the proof of sample complexity lower bound of entanglement-free schemes, we choose $2\sigma^2=0.99 \kappa$ and $R^2=\kappa n$.
For our purpose, it suffices to show that $\frac{\Gamma(n,{n}/{0.99})}{\Gamma(n)}\le0.5$.
To see this, we use the following inequality \cite{ghosh2021exponential}:
\begin{equation}\label{eq:gaussian_tail_exp}
    \frac{\Gamma(n,kn)}{\Gamma(n)} \le (ke^{1-k})^n,\quad \forall~k>1.
\end{equation}
First notice that for $k=1/0.99$ and $n=14000$, $(ke^{1-k})^n\le 0.492$. Now, for every $n<14000$, one can numerically verify $\frac{\Gamma(n,{n}/{0.99})}{\Gamma(n)}\le0.5$ (see Fig.~\ref{fig:tail}); For $n>14000$, the upper bound $(ke^{1-k})^n$ monotonically decreases with $n$, so we also have $\frac{\Gamma(n,{n}/{0.99})}{\Gamma(n)}\le0.492$. Combining these two cases completes the proof.

\bibliography{sm.bib}